\documentclass[aip,jmp,reprint]{revtex4-1}

\usepackage[latin1]{inputenc} \usepackage{graphicx} \usepackage[all]{xy}
\usepackage{mathrsfs} \usepackage{amsmath} \usepackage{amsthm}
\usepackage{amsfonts} \usepackage{amssymb} \usepackage{subfig}
\usepackage{wrapfig}

\newtheorem{teo}{Theorem}[section] \newtheorem{lemma}{Lemma}[section]
\newtheorem{prop}{Proposition}[section] \theoremstyle{definition}
\newtheorem{defin}{Definition}[section] 
\newtheorem{es}{Example}[section]

\begin{document}

\title{The extension problem for partial Boolean structures in Quantum
Mechanics} 
\author{Costantino Budroni} \affiliation{Dipartimento di Fisica, Universit\`a di Pisa, Italy} 
\author{Giovanni Morchio} \affiliation{Dipartimento di Fisica, Universit\`a di Pisa, Italy} \affiliation{INFN, Sezione di Pisa, Italy}

\begin{abstract}
Alternative \textit{partial Boolean structures}, implicit
in the discussion of classical representability of sets of 
quantum mechanical predictions, are characterized, with 
definite general conclusions on the equivalence of the 
approaches going back to Bell and Kochen-Specker.
An algebraic approach is presented, allowing for a discussion of 
\textit{partial classical extension}, amounting to  
reduction of the ``number of contexts'',  classical
representability arising as a special case. 
As a result, known techniques are generalized and 
some of the associated computational difficulties overcome. 
The implications on the discussion of Boole-Bell inequalities 
are indicated.

\end{abstract}

\maketitle

\section{Introduction}

The central role of \textit{partial classical} (Boolean) \textit{structures}
in Quantum Mechanics (QM) has been recognized by many authors, both for their
abstract mathematical implications  (constraints on \textit{truth} and
\textit{probability} assignments, Gleason theorem \cite{Glea}, Kochen-Specker
theorem \cite{K-S}) and for their fundamental role in the analysis of empirical
correlations (Bell \cite{Bell64}, Mermin \cite{Merm98}, Pitowsky
\cite{Pit89}).

Most notably, Gleason's result shows that the partial Boolean structure of QM
given by the set of Boolean algebras of projections of a
Hilbert space of dimension $\geq 3$ forces the corresponding sets of
probability measures to be given by quantum states.

Maximal Boolean algebras define \textit{contexts}, corresponding to the 
QM notion of \textit{jointly measurable  observables}; more generally, 
they play the role of maximal sets of observables for which a  classical 
description is given. Independently of QM, they arise in general as a result of 
\textit{compatibility relations} (which can be formulated \cite{B-M} 
in terms of few properties of sequences of experiments) characterizing 
sets of observables on which logical operations can be defined. 

A basic problem behind the introduction of partial 
Boolean structures is its necessity, i.e. whether their logical and 
probabilistic structures are compatible or not with a classical 
probability theory.
More generally, one may ask whether partial classical extensions exist, 
giving joint probabilistic predictions for certain sets of 
\textit{incompatible} observables.

In this language, the classical \lq\lq hidden variable\rq\rq\ problem concerns
the existence of a \textit{classical representation}, identifying 
partial Boolean structures and measures on them as restrictions of a single 
probability theory. More generally, the existence of a classical description for
sets larger than the given contexts plays an important role in the 
discussion of the interpretation of QM and gives rise to a more 
general extension problem. 

The object of the present paper is a general analysis of partial classical
structures, given by \textit{partial Boolean algebras (PBAs)}
and  \textit{partial probability theories (PPTs)} and of their extension 
problem. As we shall see, our analysis will cover rather general notions 
and relations, with substantial implications on the analysis of the interpretation 
of QM, which seem to have been overlooked.  

In Sect.\ref{sec:ppt}  the notions of PBA, PPT and \textit{extension} are introduced
and discussed.

In Sect.\ref{sec:eq} alternative PBAs and PPTs are associated to QM observables and 
predictions.
In fact, one may either consider collections of Boolean 
algebras of commuting projections, or partial Boolean algebras \textit{freely 
generated} by $yes/no$ observables and treat Boolean relations as
empirical relations, induced by \lq\lq quotients\rq\rq\ associated to 
experiments. 

We will show that such quotients are well defined for PBAs arising 
in the QM case and lead from free algebras to projection algebras.
The distinction between free abstract algebras and concrete projection
algebras is essential in order to obtain a unification 
(Theorem \ref{teo:empquoz} below) of 
Kochen-Specker-type and Bell-type approaches to the investigation of 
classical representability.

The aim of rest of the paper is to discuss the problem of the
extension of partial Boolean structures.
Classical representability is the most studied issue and is 
usually discussed in terms of Bell-like inequalities \cite{Pit89}.
Their violation in QM excludes such a representability in general; 
on the contrary, as we shall see, non-trivial partial extensions 
arise automatically in many cases; on one side this gives an interpretation of
Bell-like inequalities as conditions for further extensions and allows for
a more constructive discussion of their violation; on the other side, such 
extensions give rise to a simplification of the computation of conditions of
classical representability, reducing the problem to the compatibility of 
automatically provided solutions for certain subproblems.

In Sect.\ref{sec:34} the extension problem is discussed for all the $3$ and $4$ observables
cases which are relevant for QM.

In Sect.\ref{sec:new}  we present general results and techniques for the computation
of extensibility conditions in terms of topological properties of
compatibility relations.

In Appendix \ref{sec:ba} we recall some basic notions and results for Boolean algebras.

In Appendix \ref{sec:cp} we collect results that explicitly relate the correlation
polytope approach to our algebraic approach.

In Appendix \ref{sec:ht} we recall Horn and Tarski's notion of \textit{partial measure}
which provides, in our framework, an extensibility criterion for a large class
of partial probability theories.

\section{\label{sec:ppt}Partial probability theories}
We start introducing PBAs and PPTs. The basic notions go back to
Kochen and Specker \cite{K-S}; our approach is more general,
since we do not assume a property, indicated in the following as 
\textit{(K-S)}, which holds for Boolean structures in QM; its role in the
extension problem will be discussed below.
 
A \textit{partial Boolean algebra (PBA)} is a set $X$ together with a
non-empty family $\mathcal{F}$ of Boolean algebras,
$\mathcal{F}\equiv\{\mathfrak{B}_i\}_{i\in I}$, such that $\bigcup_i
\mathfrak{B}_i=X$, that satisfy
\begin{itemize}
 \item[$(P_1)$] for every $\mathfrak{B}_i,\mathfrak{B}_j\in \mathcal{F}$,
$\mathfrak{B}_i\cap\mathfrak{B}_j \in \mathcal{F}$ and the
Boolean operations $(\cap_i,\cup_i,^{c_i})$, $(\cap_j,\cup_j,^{c_j})$ of
$\mathfrak{B}_i$ and $\mathfrak{B}_j$ coincide on it.
\end{itemize}
Without  loss of generality we can also assume the property
\begin{itemize}
\item[$(P_2)$] for all $\mathfrak{B}_i\in \mathcal{F}$, each Boolean
subalgebra of $\mathfrak{B}_i$ belongs to $\mathcal{F}$.
\end{itemize}

By $(P_1)$, Boolean operations, when defined, are unique and will be denoted
by $(\cap, \cup, ^{c})$; we shall denote a
partial Boolean algebra by $(X,\{\mathfrak{B}_i\}_{i\in I})$, or
simply by $\{\mathfrak{B}_i\}_{i\in I}$.  In the following we
shall consider only \textit{finite} partial Boolean algebras. Their 
elements will also be called \textit{observables}. \\

Given a partial Boolean algebra $(X,\{\mathfrak{B}_i\})$, a
\textit{state} is defined as a map
$f:X\longrightarrow[0,1]$, such that $f_{|_{\mathfrak{B}_i}}$ is a normalized
measure on the Boolean algebra $\mathfrak{B}_i$ for all $i$. Equivalently, a
state is given by a collection of \textit{compatible
probability measures} $\{\mu_i\}$, i.e. measures coinciding on intersections
of Boolean algebras, one for each  $\mathfrak{B}_i$.\\

A \textit{partial probability theory (PPT)} is  a pair
$((X,\{\mathfrak{B}_i\}); f)$,  where
${(X,\{\mathfrak{B}_i\})}$ is a partial Boolean algebra and $f$
is a  state defined on it.  Equivalently, a partial probability
theory can be denoted with $((X,\{\mathfrak{B}_i\}); \{\mu_i\})$,
where $\mu_i=f_{|_{\mathfrak{B}_i}}$, or simply by $(\{\mathfrak{B}_i\};
\{\mu_i\})$.\\

It can be easily checked that the above properties are satisfied by the set of
all orthogonal projections in a Hilbert space of arbitrary dimension, with
Boolean operations defined by
\begin{equation}\label{eq:qmBoole}
P\cap Q\equiv PQ,\quad P\cup Q\equiv P+Q-PQ, \quad P^c\equiv1-P,
\end{equation}
 for all pairs $P,Q$ of commuting projections. If
one considers a finite set of projections, the result  
of the iteration of the above Boolean operations (on commuting projections) 
is still a finite set and a partial Boolean algebra.  

Moreover, given a set of projections, the corresponding
predictions given by a QM state define a PPT on the generated PBA.
In fact, given a PBA of projections 
on a Hilbert space $\mathcal{H}$, by the spectral theorem, a quantum 
mechanical state $\psi$ defines a state $f_\psi$ on it,
given by $f_\psi(P)=(\psi, P \psi)$.
The generalization to density matrices is obvious.

We shall name the so obtained PPTs
\textit{projection algebra partial probability theories}. 
We shall see in Sect. \ref{sec:eq} that they are not the only PPTs that can be 
associated to QM predictions, other choices being implicit in different
approaches to contextuality in QM.

It is interesting to notice that, in QM, PBAs of projections
also satisfy the following property
\begin{itemize}
\item[\textit{(K-S)}] if $A_1,\ldots,A_n$ are elements of $X$ such that any
two of them belong to a common algebra $\mathfrak{B}_i$, then there is a
$\mathfrak{B}_k\in \mathcal{F}$ such that $A_1,\ldots,A_n\in \mathfrak{B}_k$;
\end{itemize}
which is actually  part of the definition of partial Boolean algebra given by
Kochen and Specker.

The reason for not assuming \textit{(K-S)} is that it seems to be only motivated by
PBAs arising in QM. In a general theory of measurements, it makes
perfectly sense to consider, for instance, three measurements such that every
pair can be performed jointly, but it is impossible to perform jointly all the
three. Moreover, PPTs arising in such a case are not in general given by a 
probability on a common Boolean algebra, and therefore property \textit{K-S} 
is a real restriction. 
\newline

Given a PBA $(X,\{\mathfrak{B}_i\})$, we
shall call a \textit{context} each maximal, with respect to inclusion,
Boolean algebra of $\{\mathfrak{B}_i\}$.  Moreover, given $A,B\in X$, we shall
say that $A$ and $B$ are \textit{compatible} if they belong to a common
context.

Given a subset 
$\mathcal{G}\subset X$, we shall say that $\mathcal{G}$ \textit{generates}, 
or that $\mathcal{G}$ is a set of \textit{generators} for
$(X,\{\mathfrak{B}_i\})$, 
if each maximal Boolean algebra of $\{\mathfrak{B}_i\}$ 
is generated by a subset of $\mathcal{G}$.

Given two partial Boolean algebras $(X,\{\mathfrak{B}_i\})$ and
$(X',\{\mathfrak{B}'_j\})$; we say that a function
${\varphi:X\rightarrow X'}$ is a \textit{homomorphism} 
if for each $\mathfrak{B}_i$ the image $\varphi({\mathfrak{B}_i})$ 
belongs to $\{\mathfrak{B}'_j\}$ and 
$\varphi_{|_{\mathfrak{B}_i}}$
is a homomorphism of Boolean algebras; moreover, if $\varphi$ is
invertible, we say that $\varphi$ is an \textit{isomorphism}. 
If $(X',\{\mathfrak{B}'_j\})$ is a Boolean
algebra (notice that a Boolean algebra is also a PBA) and
the homomorphism $\varphi$ is an injection, we say that 
$\varphi$ is an \textit{embedding}.
Homomorphisms of $(X,\{\mathfrak{B}_i\})$ into the Boolean algebra 
$\{0,1\}$ define \textit{multiplicative}  states.
\\

In the following, we shall analyze the possibility of extending a partial
probability theory to additional algebras, reducing the number of contexts. 

We shall say that $(X', \{\mathfrak{B}'_j\})$
\textit{contains} $(X, \{\mathfrak{B}_i\})$  if 
$ X \subset X' $ and
$\{\mathfrak{B}_i\}\subset\{\mathfrak{B}'_j\}$.

We shall say that $(X',\{\mathfrak{B}'_j\})$ \textit{extends}
$(X, \{\mathfrak{B}_i\})$ if $(X', \{\mathfrak{B}'_j\})$ contains 
$(X, \{\mathfrak{B}_i\})$ and 
$X$ generates $(X', \{\mathfrak{B}'_j\})$.

Similar notions apply to states. Given two PPTs
$\mathcal{C}=((X, \{\mathfrak{B}_i\});\{\mu_i\})$ and
$\mathcal{C}'=((X', \{\mathfrak{B}'_j\});\{\mu'_j\})$, we shall say that 
$\mathcal{C}'$ \textit{contains} $\mathcal{C}$ if
$(X', \{\mathfrak{B}'_j\})$ contains $(X, \{\mathfrak{B}_i\})$ and
$\{\mu_i\}\subset\{\mu'_j\}$; we shall say that $\mathcal{C}$ 
\textit{extends} $\mathcal{C}'$ if $(X', \{\mathfrak{B}'_j\})$ extends
$(X, \{\mathfrak{B}_i\})$ and $\mathcal{C}'$ contains $\mathcal{C}$.

By \textit{classical representation} of a PPT 
${\mathcal{C}=((X,\{\mathfrak{B}_i\});\{\mu_i\})}$ 
we shall mean a Boolean algebra $\mathfrak{B}$ and a (normalized) measure $\mu$ 
such that $(\mathfrak{B}; \mu)$  extends ${\mathcal{C}}$. 

The fact that a PBA is not embeddable into a Boolean algebra 
is precisely the original form of the Kochen-Specker theorem.
The minimality implicit in the above notion of extension 
reduces the multiplicity of classical representations in the 
sense of Kochen and Specker \cite{K-S} (not requiring that the PBA 
\textit{generates} the Boolean algebra); 
however, a classical representation
exists in our sense iff it exists in the K-S sense since 
clearly a PBA is embeddable in a Boolean algebra iff it 
can be extended to a Boolean algebra.

If its PBA $\{\mathfrak{B}_i\}$  extends to
a Boolean algebra $\mathfrak{B}$, the existence of a classical
representation of a PPT amounts to the extension problem of a function, 
induced by the corresponding state, defined on a subset 
of $\mathfrak{B}$; the solution of this extension problem (with 
necessary and sufficient conditions) is then implicit in the work 
of Horn and Tarski \cite{Tarski}, which is summarized in Appendix \ref{sec:ht}. 
A PPT $\mathcal{C}=(\{\mathfrak{B}_i\};\{\mu_i\})$ 
such that $\{\mathfrak{B}_i\}$ 
extends to a Boolean algebra will be called a 
\textit{Horn-Tarski (H-T) partial probability theory}.

\section{\label{sec:eq}Reduction to Horn-Tarski PPTs}

\subsection{Empirical quotients of partial probability theories}

The aim of the following discussion is to show how PBAs and PPTs 
provide a unification of the Kochen-Specker-type and Bell-type 
approaches to classical representability.

A fundamental role is played by the notion of \textit{empirical quotient}; 
we shall  briefly discuss it in classical probability theory and then 
we shall generalize it to PPTs.

Consider a classical probability theory defined by a finite Boolean algebra 
$\mathfrak{B}$ and a probability measure $\mu$. 
If for two elements $A,B\in\mathfrak{B}$ it holds $\mu(A\cap B^c)=\mu(A^c\cap
B)=0$, equivalently $\mu(A)=\mu(B)=\mu(A\cap B)$, it follows that every time 
$A$ happens also $B$ happens and conversely. In terms of conditional
probabilities this can be written as $Pr(A | B)=Pr(B | A)=1$. Therefore, in
the situations described by the measure $\mu$, it makes sense
to identify the events $A,B$ and $A\cap B$ with a single event since 
they \lq\lq cannot be distinguished by any experiment\rq\rq.

This procedure 
induces an equivalence relation $\sim_{\mathcal{I}}$  on $\mathfrak{B}$, given
by the ideal $\mathcal{I}=\{A\in\mathfrak{B}|\mu(A)=0\}$, giving rise to 
the \textit{empirical quotient algebra}
$\widetilde{\mathfrak{B}}\equiv \mathfrak{B}/_{\sim_{\mathcal{I}}}$.
$\mu$ induces a normalized measure $\tilde{\mu}$ on
 $\widetilde{\mathfrak{B}}$.

Similar notions, with identical interpretation, apply to the case of a finite Boolean
algebra $\mathfrak{B}$ 
and a collection of normalized measures
$\{\mu_k\}_{k\in K}$, where $K$ may be any set
of indices, through the ideal
$\mathcal{I}=\{A\in\mathfrak{B}|\mu_k(A)=0$ for all $k\in K\}$
(any K being admissible since $\mathfrak{B}$ is finite). \\

The extension of the above notions to the case of PPTs is not
automatic and requires further conditions. 

Given two collections of PPTs
$\{\mathcal{C}_k\}_{k\in K}=\{(\{\mathfrak{B}_i\}_{i\in I};f_k)\}_{k\in K}$
and \\$\{\widetilde{\mathcal{C}}_k\}_{k\in K}=
\{(\{\widetilde{\mathfrak{B}}_j\}_{j\in J};\tilde{f}_k)\}_{k\in K}$,
we shall say that  $\{\widetilde{\mathcal{C}}_k\}_{k\in K}$ is an 
\textit{empirical quotient} of $\{\mathcal{C}_k\}_{k\in K}$ if 
there exists an \textit{equivalence relation} $\sim$ on 
$X=\bigcup_i \mathfrak{B}_i$ such that
\begin{itemize}

\item[$(i)$] when restricted to each Boolean algebra 
$\mathfrak{B}_i$, $\sim$ coincides with the equivalence relation induced 
by the ideal $\mathcal{I}_i\equiv\{ A\in\mathfrak{B}_i | f_k(A)=0 
\text{ for all } k\in K\}$;

\item[$(ii)$] given $A\in \mathfrak{B}_i$ and $B\in\mathfrak{B}_l$, with
 $\mathfrak{B}_i$ and $\mathfrak{B}_l$ maximal, if $A\sim B$, then there 
 exists $C\in\mathfrak{B}_i\cap\mathfrak{B}_l$ such that $A\sim C$ 
 (and $B\sim C$ by transitivity);

\item[$(iii)$]  the quotient set $X/_{\sim}$ is a partial Boolean algebra
isomorphic to the PBA $\widetilde{X}=\bigcup_j\widetilde{\mathfrak{B}}_j$;
by $(i)$, this implies that the quotient preserves Boolean operations, namely
for all $A,B\in X$, with $A$ and $B$ compatible, it holds 
$[A]\cap [B]=[A\cap B]$, where $[A]$ denotes the equivalence class of 
$A$ with respect to $\sim$, and analogous properties hold for $\cup$ 
and $^c$;

\item[$(iv)$] denoted with $\varphi : X/_{\sim} \longrightarrow \widetilde{X}$ 
the isomorphism in $(iii)$, it holds $f_k(A)= \tilde{f}_k(\varphi([A]))$, 
for all $k\in K$ and for all $A\in X$.

\end{itemize}

The above definition clearly applies in the classical case,
 i.e. when both $X$ and $\widetilde{X}$ are Boolean algebras; 
 we shall provide below less trivial examples.
 
We remark that, unlike the classical case, an equivalence relation on a PPT
satisfying (i) and (iv)  does not in general give rise to an empirical quotient;
a counterexample can be constructed by considering a PPT given by the PBA 
consisting of three maximal Boolean algebras, generated respectively by the 
pairs of observables $\{A,B\}$, $\{B,C\}$ and $\{A,C\}$, together with 
the corresponding subalgebras, and a state $f$ 
that induces in the above Boolean algebras the identification 
$A\sim B$, $B\sim C$ and $C\sim A^c$.

In fact, if an empirical quotient exists, then by transitivity
$A$ is identified with $A^c$ and therefore, by $(i)$, both are 
identified with $\emptyset$; 
this contradicts $\tilde{f}(\varphi([\mathbf{1}]))=1$. 

The above notion of quotient may look too restrictive; 
on the contrary, it will turn out that \textit{all} PPTs
with a PBA admitting a complete set of states (see below) 
can be \textit{identified with quotients} of PPTs associated to a 
collection of freely generated Boolean algebras, automatically 
embeddable into a Boolean algebra.
This will imply that \textit{all extension problems} in QM can 
be put in the H-T form.

\subsection{Classical representations of partial probability 
theories and of their empirical quotients} 

An important role is played by the following notions.

Given a PBA $\{\mathfrak{B}_i\}_{i\in I}$ and a collection
of  states $\{f_k\}_{k\in K}$, we shall say that the collection
$\{f_k\}_{k\in K}$ is \textit{complete} with respect to
$\{\mathfrak{B}_i\}_{i\in I}$ if for all $A\in X=\bigcup_i \mathfrak{B}_i$,
with $A\neq \emptyset$ there exists $f_k$ such that $f_k(A)\neq 0$. If, in addition,
for all $A\neq B$, with $A,B\in X$, there exists $f_k$ such that $f_k(A)\neq f_k(B)$
then $\{f_k\}_{k\in K}$ is said to be \textit{separating} for
$\{\mathfrak{B}_i\}_{i\in I}$.

Notice that for an empirical quotient
$\{\widetilde{\mathcal{C}}_k\}_{k\in K}
=\{(\{\widetilde{\mathfrak{B}}_j\}_{j\in J};\tilde{f}_k)\}_{k\in K}$,
by $(i)$ and $(iv)$, $\{\tilde{f}_k\}_{k\in K}$ 
is always complete with respect 
to $\{\widetilde{\mathfrak{B}}_j\}_{j\in J}$.

The following result relates classical representations of PPTs
with embeddings of PBAs associated to empirical quotients.

\begin{prop}\label{prop:01mis}
 Given ${\{\mathcal{C}_k\}_{k\in K}=}$\\ ${\{(\{\mathfrak{B}_i\}_{i\in I};f_k)\}_{k\in
K}}$ and ${\{\widetilde{\mathcal{C}}_k\}_{k\in
K}=\{(\{\widetilde{\mathfrak{B}}_j\}_{j\in J};\tilde{f}_k)\}_{k\in K}}$, with
$\{\widetilde{\mathcal{C}}_k\}_{k\in K}$ an empirical quotient of
$\{\mathcal{C}_k\}_{k\in K}$, if there exists $k_0\in K$ such that
${\mathcal{C}}_{k_0}$ admits a classical representation, then there exists a
multiplicative state on $\{\widetilde{\mathfrak{B}}_i\}$, i.e. a
homomorphism $\delta_{0}:\widetilde{X}=
\bigcup_j\widetilde{\mathfrak{B}}_j\longrightarrow \{0,1\}$.

Moreover, if there exists $K'\subset K$ such that $\{\tilde{f}_k\}_{k\in K'}$ is 
separating for $\{\widetilde{\mathfrak{B}}_j\}_{j\in J}$ and
${\mathcal{C}}_{k}$ admits a classical representation for  every $k\in K'$,
then  $\{\widetilde{\mathfrak{B}}_j\}_{j\in J}$ is
embeddable into the Boolean algebra $2^N$, the power set of a $N$-element set,
where $N$ is the number of multiplicative  states induced by 
classical representations of the states $\{f_k\}_{k\in K'}$.
\end{prop}

\textbf{Proof} Let the Boolean algebra $\mathfrak{B}$ together with the
normalized measure $\mu$ be a classical representation for
$\mathcal{C}_{k_0}$, then $\mu$ can be written as a convex combination of
multiplicative measures (see Lemma \ref{lemma:mismolt} below), namely
\begin{equation} \label{eq:muconv}
\mu=\sum_i\lambda_i\delta_i,
 \end{equation}
 where the $\delta_i$'s are multiplicative measures and the $\lambda_i$'s are
positive numbers that sum up to one. It  follows that $\mu(A\cap
B^c)=\mu(A^c\cap B)=0$ for all $A,B\in X$ such that $A\sim B$ and $A$ and $B$
belong to a common algebra $\mathfrak{B}_{i_0}\in\{\mathfrak{B}_i\}$; 
therefore $\delta_i(A\cap B^c)=\delta_i(A^c\cap B)=0$
for each $\delta_i$ that appears in (\ref{eq:muconv}). Actually, 
the same holds even if $A$ and $B$ do not belong to a common maximal
algebra of $\{\mathfrak{B}_i\}$. In fact, by $(ii)$, there
exists an element $C$ in the intersection of the two maximal algebras
containing $A$ and $B$ such that $A\sim C\sim B$ and the above statement
follows from  ${A\cap B^c=(A\cap B^c\cap C)\cup (A\cap B^c\cap C^c)}$.

It follows that $\delta_i(A)=\delta_i(B)$ for all
$A,B\in X$ such that $A\sim B$ and for all $\delta_i$ appearing in
(\ref{eq:muconv}); therefore each $\delta_i$ induces a well defined
$\{0,1\}$-valued function on $\widetilde{X}$. To conclude, we shall prove that
such functions are homomorphisms when restricted to each algebra of
$\{\widetilde{\mathfrak{B}}_j\}$. This follows from the isomorphism between
 $\widetilde{X}$  and $X/_{\sim}$ and the fact
that each $\delta_i$ defines a multiplicative measure on
$\mathfrak{B}_i/_{\sim}$ for all $\mathfrak{B}_i$. In fact, given $A,B\in
\mathfrak{B}_i$, $[A]\cap [B]=[\emptyset]$ implies $\delta_i(A\cap B)=0$ and
therefore $\delta_i(A)+\delta_i(B)=\delta_i(A\cup B)$; each $\delta_i$ defines,
therefore, a $\{0,1\}$-valued function on $\mathfrak{B}_i/_{\sim}$ which is
additive on disjoint elements, i.e. a multiplicative measure, which is a
homomorphism with the Boolean algebra $\{0,1\}$ (see Lemma
\ref{lemma:atomlib}).

The proof of the second part follows easily from the first part together with
Theorem 0 of Ref. \onlinecite{K-S}. $\square$\\

\subsection{\label{subs:empq}Partial probability theories as empirical 
quotients of free H-T theories}

We now show that any complete set of states on a PBA can be regarded as
an empirical quotient of a collection of PPTs on a PBA which is embeddable in
a (free) Boolean algebra, i.e a collection of H-T PPTs.

Consider a collection of PPTs
$\{\widetilde{\mathcal{C}}_k\}_{k\in
K}=\{(\{\widetilde{\mathfrak{B}}_j\}_{j\in J};\tilde{f}_k)\}_{k\in K}$ such
that $\{\tilde{f}_k\}_{k\in K}$ is complete, and take a subset
$\widetilde{\mathcal{G}}=\{\widetilde{A}_1,\ldots,\widetilde{A}_n\}\subset\widetilde{X}=\bigcup_j
\widetilde{\mathfrak{B}}_j$ of generators of
$\{\widetilde{\mathfrak{B}}_j\}_{j\in J}$ satisfying the following property
\begin{itemize}
 \item[$(G)$] given $k\geq 1$ maximal Boolean algebras
 $\widetilde{\mathfrak{B}}_{i_1},\ldots,\widetilde{\mathfrak{B}}_{i_k}$,
 generated respectively by maximal subsets of compatible generators
 ${\widetilde{\mathcal{G}}_{i_1},\ldots,\widetilde{\mathcal{G}}_{i_k}\subset\widetilde{\mathcal{G}}}$,
 such that
 $\widetilde{\mathfrak{B}}_{i_1}\cap\ldots\cap\widetilde{\mathfrak{B}}_{i_k}\neq
 \{\emptyset,\mathbf{1}\}$, the set
 $\widetilde{\mathcal{G}}_{{i_1}\ldots{i_k}}\equiv\widetilde{\mathcal{G}}_{i_1}\cap\ldots\cap\widetilde{\mathcal{G}}_{i_k}$
 is not empty and it generates the Boolean algebra
 $\widetilde{\mathfrak{B}}_{i_1}\cap\ldots\cap\widetilde{\mathfrak{B}}_{i_k}$;
\end{itemize}
  notice that each maximal algebra is generated by a maximal subset of
  compatible generators and that the above choice is always possible since one
  can take $\widetilde{\mathcal{G}}=\widetilde{X}$. The role of this property
  will be clarified below.
  
  Denote with $\{\widetilde{\mathcal{G}}_l\}$ the collection of subsets of
 compatible observables of $\widetilde{\mathcal{G}}$,
$\widetilde{\mathcal{G}}_l = \{ \tilde A_{s_1} \ldots \tilde A_{s_{n_l}} \}$.  
 Now consider the
 PBA $\{\mathfrak{B}_i\}_{i\in I}$ consisting of 
 Boolean algebras freely generated by subsets 
$\mathcal{G}_l \equiv \{ A_{s_1} \ldots A_{s_{n_l}} \}$.   

We now show how each state $\tilde{f}_k$ induces a  state
$f_k$ on $\{\mathfrak{B}_i\}_{i\in I}$. First, notice that, since each
 state on a PBA is a collection of normalized measures, it is sufficient to
define it as measures on maximal Boolean algebras. Each measure on a maximal
algebra  $\mathfrak{B}_l$ of $\{\mathfrak{B}_i\}_{i\in I}$, generated by a set
$\mathcal{G}_l=\{A_{s_1},\ldots, A_{s_{n_l}}\}$, is completely determined by
its values on elements of the form $(-1)^{1-\varepsilon_1}
A_{s_1}\cap\ldots\cap (-1)^{1-\varepsilon_{n_l}} A_{s_{n_l}}$, where $-A\equiv A^c$
and $\varepsilon_i\in\{0,1\}$, since each element of the algebra can be
written as a disjoint union of elements of that form (see Lemmas
\ref{lemma:mismolt} and \ref{lemma:atomlib}).  Now, $f_k$ is defined as $f_k((-1)^{1-\varepsilon_1}
A_{s_1}\cap\ldots\cap (-1)^{1-\varepsilon_{n_l}} A_{s_{n_l}})\equiv
\tilde{f}_k((-1)^{1-\varepsilon_1} \widetilde{A}_{s_1}\cap\ldots\cap
(-1)^{1-\varepsilon_{n_l}} \widetilde{A}_{s_{n_l}})$ for all maximal subsets of compatible
observables $\widetilde{\mathcal{G}}_l$ of $\widetilde{\mathcal{G}}$, and
extended as a measure on each maximal algebra. It can be verified that such
measures are normalized and they coincide on intersection of Boolean algebras;
therefore, they define a state.\\

In this way, we obtain a collection of PPTs
$\{\mathcal{C}_k\}_{k\in K}\equiv\{(\{\mathfrak{B}_i\}_{i\in I};f_k)\}_{k\in
K}$ such that the initial collection $\{\widetilde{\mathcal{C}}_k\}_{k\in
K}=\{(\{\widetilde{\mathfrak{B}}_j\}_{j\in J};\tilde{f}_k)\}_{k\in K}$ is an
empirical quotient. 
The equivalence relation $\sim$ can be, in fact, defined
as follows: to each element $A$ of $X$, generated by a subset of compatible
generators $\mathcal{G}_l\subset \mathcal{G}$ there corresponds, via the 
correspondence  $A_i\mapsto \widetilde{A}_i$, a unique element $\widetilde{A}$ 
of $\widetilde{X}$, defined as the element generated by 
$\widetilde{\mathcal{G}}_l\subset\widetilde{\mathcal{G}}$
by means of the same operations that generate $A$ from   $\mathcal{G}_l$; then 
 an equivalence relation $\sim$ can be defined on $X$ as $A\sim B$ iff 
 $\widetilde{A}=\widetilde{B}$.

It can be easily verified that $\sim$ is an equivalence relation and that
it defines an empirical quotient:

$(i)$: it is sufficient to consider each Boolean algebras $\mathfrak{B}_l$,
generated by $\mathcal{G}_l=\{A_{l_1},\ldots,A_{l_s}\}$, and  notice that, there,
$\sim$ coincides with the equivalence relation induced by the ideal 
${\mathcal{I}\equiv\{B\in\mathfrak{B}_l |
\bigcup_{\varepsilon\in H_B}(-1)^{1-\varepsilon_1}\widetilde{A}_{l_1}\cap\ldots\cap
(-1)^{1-\varepsilon_{s}}\widetilde{A}_{l_s}=\emptyset\}}$ with
$H_B\equiv\{\varepsilon=(\varepsilon_1,\ldots,\varepsilon_n)\in\{0,1\}^n|
(-1)^{1-\varepsilon_1}A_{l_1}\cap\ldots\cap (-1)^{1-\varepsilon_s}A_{l_s}\subset B\}$
(see lemma \ref{lemma:atomlib} below); 
now, since $\{\tilde{f}_k\}_{k\in K}$ is complete
and by construction of $\{f_k\}_{k\in K}$,
$\mathcal{I}$ coincides with the set 
$\{B\in\mathfrak{B}_l | f_k(B)=0$ for all $k\in K\}$.

$(ii)$: given $A,B\in X$, belonging respectively to maximal algebras 
$\mathfrak{B}_{l_1}$, generated by $\mathcal{G}_{l_1}$, and 
$\mathfrak{B}_{l_2}$, generated by $\mathcal{G}_{l_2}$, with 
$\mathcal{G}_{l_1}$ and $\mathcal{G}_{l_2}$ maximal, if $A\sim B$,
then there exists $C\in \mathfrak{B}_{l_1}\cap \mathfrak{B}_{l_1}$, which is
the Boolean algebra generated by $\mathcal{G}_{l_1}\cap \mathcal{G}_{l_2}$,
such that $A\sim C\sim B$. In fact, $A\sim B$ implies, with the same notation
as above, $\widetilde{A}=\widetilde{B}$; therefore the two maximal algebras
generated respectively by  $\widetilde{\mathcal{G}}_{l_1}$ and 
$\widetilde{\mathcal{G}}_{l_2}$ have a non-empty intersection containing 
$\widetilde{A}$, then, by $(G)$,
 $\mathcal{G}_{l_1}\cap \mathcal{G}_{l_2}\neq \emptyset$ and an element 
 $C$ satisfying the above conditions exists.

$(iii)$: by construction, 
$X/_\sim$ is in a one-to-one correspondence with $\widetilde{X}$;
that such a bijection is also an isomorphism follows from the coincidence, within each
Boolean algebra, of $\sim$ with the equivalence relation induced by the ideal 
$\mathcal{I}$ discussed above.

$(iv)$: it follows by construction of $\{f_k\}_{k\in K}$.\\

The above partial Boolean algebra $\{\mathfrak{B}_i\}_{i\in I}$ 
is embeddable into the Boolean algebra freely generated by the set 
$\mathcal{G}$. The PPTs $\{\mathcal{C}_k\}_{k\in K}$ are therefore of the 
Horn-Tarski type and we shall name 
$\{\mathcal{C}_k\}_{k\in K}$ \textit{the collection of free H-T
 partial probability theories associated to}
 $\{\widetilde{\mathcal{C}}_k\}_{k\in K}$ \textit{and}
 $\widetilde{\mathcal{G}}$.

\subsection{Classical representations and free H-T theories}
The following theorem applies the results of Proposition \ref{prop:01mis}
to the above construction, allowing to reduce the discussion of 
the existence of classical representations to H-T theories.

\begin{teo}\label{teo:empquoz} 
 Given a collection of PPTs
 $\{\widetilde{\mathcal{C}}_k\}_{k\in
 K}=\{(\{\widetilde{\mathfrak{B}}_j\}_{j\in  J};\tilde{f}_k)\}_{k\in K}$ with
 $\{\tilde{f}_k\}_{k\in K}$ complete with respect to
 $\{\widetilde{\mathfrak{B}}_j\}_{j\in  J}$, a set of generators
 $\widetilde{\mathcal{G}}=\{\tilde{A}_1,\ldots,\tilde{A}_n\}$ satisfying
 property $(G)$ and the associated collection  of free H-T PPTs
 $\{\mathcal{C}_k\}_{k\in K}=\{(\{\mathfrak{B}_i\}_{i\in I};f_k)\}_{k\in K}$,
 then
 \begin{itemize}
  \item[$(a)$] if, for a given $k\in K$,  $\widetilde{\mathcal{C}}_k$ admits a
  classical representation, then ${\mathcal{C}}_k$  admits a classical
  representation;
  
 \item[$(b)$] if there exists $K'\subset K$ 
such that $\{\tilde{f}_k\}_{k\in K'}$ is separating for 
 $\{\widetilde{\mathfrak{B}}_j\}_{j\in  J}$ and   ${\mathcal{C}}_{k}$ admits a
 classical representation for all $k\in K'$, then $\widetilde{\mathcal{C}}_k$
 admits a  classical representation for all $k\in K'$.
 \end{itemize}
\end{teo}
\textbf{Proof} $(a)$ Let the Boolean algebra $\widetilde{\mathfrak{B}}$
together with the normalized measure $\tilde{\mu}_k$ be a classical
representation for $\widetilde{\mathcal{C}}_k$. By the definition of extension,
 the set $\widetilde{\mathcal{G}}$ is a set of generators for
$\widetilde{\mathfrak{B}}$; therefore the Boolean algebra
$\widetilde{\mathfrak{B}}$ is isomorphic to the quotient algebra
$\mathfrak{B}/_\sim$, where $\mathfrak{B}$ is the Boolean algebra freely
generated by $n$ generators $\{A_1,\ldots,A_n\}$ and the equivalence relation
$\sim$ is that induced by the ideal ${\mathcal{I}\equiv\{B\in\mathfrak{B} |
\bigcup_{\varepsilon\in H_B}(-1)^{1-\varepsilon_1}\widetilde{A}_1\cap\ldots\cap
(-1)^{1-\varepsilon_n}\widetilde{A}_n=\emptyset\}}$ with
$H_B\equiv\{\varepsilon=(\varepsilon_1,\ldots,\varepsilon_n)\in\{0,1\}^n|
(-1)^{1-\varepsilon_1}A_1\cap\ldots\cap (-1)^{1-\varepsilon_n}A_n\subset B\}$
(see Lemma \ref{lemma:atomlib} below). Then, denoted with $\varphi$ the
isomorphism between $\mathfrak{B}/_\sim$ and $\widetilde{\mathfrak{B}}$, a
measure $\mu_k$ extending the state $f_k$ on $\mathfrak{B}$ can be
defined as $\mu_k(A)\equiv\tilde{\mu}_k(\varphi([A]))$ for all
$A\in\mathfrak{B}$, where $[A]$ is the equivalence class of $A$ with respect
to $\sim$. It can be
easily verified that $(\mathfrak{B};\mu_k)$ is a classical representation for
$\mathcal{C}_k$.

$(b)$ Let the free Boolean algebra $\mathfrak{B}$, defined as above, together
with a normalized measure $\mu_k$ be a classical representation for
$\mathcal{C}_k$, for all $k\in K'$. By Proposition \ref{prop:01mis},
$\{\widetilde{\mathfrak{B}}_j\}_{j\in J}$ is embeddable into the Boolean
algebra $2^N$, N as in Proposition \ref{prop:01mis}; 
let us denote with $\widetilde{\mathfrak{B}}$ the subalgebra of $2^N$
generated by $\widetilde{\mathcal{G}}$ and with $S$ the set of all
homomorphism $\delta:\widetilde{X}\longrightarrow \{0,1\}$ induced by the
normalized measures $\mu_k$, $k\in K'$ (see Proposition \ref{prop:01mis}). Such
homomorphisms are, by construction (see Theorem 0 in Ref. \onlinecite{K-S}), in a
 one-to-one correspondence with the multiplicative measures of $2^N$ and can be
extended to multiplicative measures on $\widetilde{\mathfrak{B}}$ in a way
uniquely determined by the values assumed on the set of generators
$\widetilde{\mathcal{G}}$. It follows that each element
$\bigcup_{\varepsilon\in H}(-1)^{1-\varepsilon_1}\widetilde{A}_1\cap\ldots\cap
(-1)^{1-\varepsilon_n}\widetilde{A}_n$, with $H\subset \{0,1\}^n$, generated by
$\widetilde{\mathcal{G}}$ is the zero element if and only if $\sum_{\varepsilon\in H}
\prod_{i=1}^n \varepsilon_i
\delta(\widetilde{A}_i)+(1-\varepsilon_i)(1-\delta(\widetilde{A}_i))=0$, i.e. the
extension of $\delta$ is zero on such an element, for all $\delta\in S$. Since
the homomorphisms in $S$ are induced by multiplicative measures associated, 
eq. (\ref{eq:muconv}), to the normalized measures $\mu_k$, $k\in K'$, it
follows that the ideal $\mathcal{I}$ defined as in $(a)$ coincides with
the ideal $\mathcal{I}'\equiv\{B\in \mathfrak{B} | \mu_k(B)=0$ for all $k\in
K'\}$. This implies, as in the proof of
Proposition \ref{prop:01mis}, that $\mu_k$ induces a normalized measure on
$\mathfrak{B}/_{\sim}$ , and consequently a normalized measure $\tilde{\mu}_k$
on $\widetilde{\mathfrak{B}}$, for all $k\in K'$. It can be easily checked that
$(\widetilde{\mathfrak{B}};\tilde{\mu}_k)$ is a classical representation for
$\widetilde{\mathcal{C}}_k$ for all $k\in K'$. $\square$\\

\subsection{Free H-T PPT versus projection algebra PPT in QM}

On the basis of the above discussion, it is clear that  
the projection algebra  
is not the only possible PBA for the formulation of QM predictions.

In particular, for any given  PBA of projections, quantum states generate
a complete collection of states on such a PBA; it follows that, 
for any set of generators satisfying property $(G)$, the construction
in Sect.\ref{subs:empq} applies and therefore 
\textit{any collection of QM predictions can be described by a free H-T PPT}. 
 
The above results formalize constructions which are often used 
implicitly in the discussion of the interpretation of QM:
consider in fact a finite set of $yes/no$ apparatuses
 $\mathcal{G}=\{A_1,\ldots,A_n\}$, represented as projections
 $\mathcal{P}=\{P_1,\ldots,P_n\}$ in a finite-dimensional Hilbert space
 $\mathcal{H}$; then for every subset of compatible apparatuses, 
i.e. commuting projections, in $\mathcal{G}$,
 it makes sense to consider logical
 combinations obtained by means of logic gates applied to the outcomes in an
 experiment where they are jointly measured. 
In this way we obtain  a
 collection of Boolean algebra of observables $\{\mathfrak{B}_i\}$, each one
 \textit{freely generated} by a subset $\mathcal{G}_i\subset\mathcal{G}$ of
 compatible apparatuses; there is no longer a bijection between
 $\{\mathfrak{B}_i\}$ and the partial Boolean algebra generated by
 $\mathcal{P}$, see eq. $(\ref{eq:qmBoole})$, but states on the PBA are
 still given by quantum mechanical states $\psi$ by
 $f_\psi(A)=(\psi,P_A\psi)$, where $A$ belong to a free Boolean algebra
 $\mathfrak{B}_i$ generated by a subset of compatible apparatuses
 $\mathcal{G}_i\subset \mathcal{G}$ and $P_A$ is the projection obtained from
 the corresponding subset of commuting projection $\mathcal{P}_i\subset
 \mathcal{P}$ by means of the same Boolean operations that generate $A$ from
 $\mathcal{G}_i$ (notice that $P_A$ may be $0$ even if $A\neq\emptyset$). 
 
Notice that the above construction only relies on the notion of 
\textit{compatible apparatuses} and \textit{observed frequencies}. 
It can be qualified as a \textit{Bell-type approach}:
every  attribution of $0$ and $1$ to a set of observables is assumed 
to be possible, only a posteriori constrained by experimental information, and
the logical structure is that of a free Boolean algebra.

A systematic treatment of such a problem is given
by Pitowsky\cite{Pit89} in terms of propositional logic; we shall
refer to it as the \textit{correlation polytope} approach
(see Appendix \ref{sec:cp} for an account in terms free Boolean algebras).\\ 
 
The alternative approach based on projection algebras gives rise
to results of a rather different form, starting from the K-S theorem,
and will be referred to as \textit{Kochen-Specker-type} approach. 

The relation between the two approaches has not been clarified in general
and is also confused by the fact that in some cases (e.g. the 
Bell argument with four measurements) the approaches seem to coincide. 

From the above discussion, it is clear the main difference between 
Kochen-Specker-type and Bell-type approaches resides in which logical 
relations between observables are assumed.

In fact, the above results imply that the K-S approach is 
related to the Bell approach by an empirical quotient:
by the construction of Sect. \ref{subs:empq}, a free H-T PPT is 
obtained from the projection algebra PPT on the basis of
any set $\mathcal{P}$  satisfying property $(G)$, which  always exists, 
as discussed above.

The logical content of such a constructions is that 
\textit{logical relations} between compatible observables 
can be weakened to 
\textit{empirical relations}, associated in principle to a 
collection of experiments or states on a PBA. 
(Similar distinctions have been introduced, with a different interpretation, 
by Garola and Solombrino\cite{Gar}).\\

The construction in Sect.\ref{subs:empq}, Proposition \ref{prop:01mis} and  
Theorem \ref{teo:empquoz} clarify 
the relation between Kochen-Specker-type results, 
presenting a non-embeddable partial Boolean algebra of projections, 
and Bell-type arguments,
giving  conditions for the existence of a probability measure
reproducing measurable correlations on a free Boolean algebra.
The result is that the equivalence of the two viewpoints 
for the discussion of classical representability in QM,
recognized by Cabello\cite{Cab08} (see also Ref. \onlinecite{BBCP09}) in 
situations arising in the discussion of the Kochen-Specker theorem,
is a very general fact, following from basic logical and probabilistic 
structures. 

In fact, Proposition \ref{prop:01mis} implies that a set of predictions that generates
a Kochen-Specker-type contradiction, namely the impossibility of a consistent truth
assignment (i.e. a homomorphism between projections PBA and $\{0,1\}$), 
also generates a  Bell-type contradiction for all quantum states 
in the associated free H-T PPTs, more precisely each quantum state 
violates at least one Bell inequality (not necessarily the same for all states).

Moreover, as a consequence of Theorem \ref{teo:empquoz}, we obtain 
that, given a set of apparatuses and a set of quantum states inducing a 
separating collection of states on their projection PBA,  
a classical representation of all the corresponding projection algebra PPTs 
exists \textit{if and only if} all the corresponding free H-T PPTs,
constructed as in Sect. \ref{subs:empq} admit a classical representation, independently of
the choice of the generators.  \\

It follows that all extension  problems arising in QM can be discussed 
in the framework of free H-T PPTs; the rest of this paper is devoted 
to the investigation of extensibility conditions in this case.

\section{\label{sec:34}Systems of $3$ and $4$ observables}
In this section we shall discuss two applications of the criterion of classical
representability, presented in Appendix \ref{sec:cp}, obtained from the translation
 of Pitowsky's correlation polytopes results into the Boolean framework.

The proofs of the following theorems are essentially
based on the analysis of Bell-Wigner and Clauser-Horne correlation polytopes
made by Pitowsky \cite{Pit89}. Theorem \ref{teo:est3} shows that for
three observables with two compatible pairs a classical
probabilistic model which reproduces observable correlations always exists;  it implies that
for three quantum mechanical observables a classical probabilistic model
always exists for all possible compatibility relations. Theorem \ref{teo:est4}
shows that for four observables with Bell-type compatibility relations a
probabilistic model for the four observables exists if and only if there are
two models for three observables that coincide on the intersection; a result
obtained by Fine \cite{Fine} in a rather different setting; our approach
provides in this case a complete analysis for the case of four quantum
mechanical observables.

\begin{teo}\label{teo:est3}
 Let $\mathfrak{B}$ be a Boolean algebra freely generated by
 $\mathcal{G}=\{A_1,A_2,A_3\}$ and $\mathfrak{B}_{13}$ and $\mathfrak{B}_{23}$
 the subalgebras generated respectively by $\{A_1,A_3\}$ and $\{A_2,A_3\}$.

 Consider $f:\mathfrak{B}_{13}\cup\mathfrak{B}_{23}\longrightarrow[0,1]$, such
that $f_{|_{\mathfrak{B}_{13}}}$ and  $f_{|_{\mathfrak{B}_{23}}}$ are
normalized measures on such subalgebras. Then $f$ is extensible to a
normalized measure on the algebra $\mathfrak{B}$.
\end{teo}
\textbf{Proof} By Lemma \ref{lemma:misunic}, without loss of generality we can
consider\\ ${f:X=\{A_1,A_2,A_3,A_1\cap A_3,A_2\cap A_3\}\longrightarrow[0,1]}$
. The vector $p=(p_1,p_2,p_3,p_{13},p_{23})$ is given by  $p_i=f(A_i)$ and
$p_{ij}=f(A_i\cap A_j)$; since such values come from a measure on
$\mathfrak{B}_{13}$ and $\mathfrak{B}_{23}$
\begin{equation}\label{eq:sub1}
 p_{13}\leq min\{p_1,p_3\}\qquad p_{23}\leq min\{p_2,p_3\};
\end{equation}
  from
\begin{eqnarray*}
&0\leq f((A_i\cup A_j)^c)=1-f(A_i\cup A_j)=\\
&=1-f(A_i)-f(A_j)+f(A_i\cap A_j),
\{i,j\}=\{1,3\},\{2,3\}
\end{eqnarray*} 
we obtain
\begin{equation}\label{eq:sub2}
 p_1 + p_3 - p_{13}\leq 1,\qquad p_2 + p_3 -p_{23}\leq 1 \ .
\end{equation}
From Lemma \ref{lemma:normmeas} and Proposition \ref{prop:polit} we know that
if a normalized measure $\mu$ which extends $f$ exists, then
\begin{equation}\label{eq:lambda} 
\lambda(\varepsilon) \equiv \mu(a_\varepsilon)=
\mu((-1)^{1-\varepsilon_1}A_1
\cap(-1)^{1-\varepsilon_2}A_2\cap(-1)^{1-\varepsilon_3}A_3)\ .
 \end{equation}
 Therefore the coefficients $\lambda(\varepsilon)$ are obtained from
(\ref{eq:lambda}) and the property $(b)$ of the  definition of measure (see
Appendix \ref{sec:ba}). The convex combination is obtained by means of two coefficients
$\chi$ and $\eta$ representing the two missing correlations $\mu(A_1\cap
A_2\cap A_3^c)$ and $\mu(A_1\cap A_2\cap A_3)$ (alternatively, one can use
$\mu(A_1\cap A_2)$ and $\mu(A_1\cap A_2\cap A_3)$, but the inequalities
$(\ref{eq:eta1})-(\ref{eq:chi2})$ below become more complicated). The
following inequalities are obtained from the non-negativity of the measure in
the same way as in $(\ref{eq:sub2})$
\begin{align}\label{eq:eta1}
 \eta\leq & min\{p_{13},\quad p_{23},\}\, \\
\label{eq:eta2}
\eta\geq & max\{0,\quad p_{13}+p_{23}-p_3\}\, \\
 \label{eq:chi1}\chi \leq & min\{ p_1-p_{13},\quad p_2-p_{23}\}\, \\
\label{eq:chi2}\chi\geq & max\{0,\quad p_1+p_2+p_3-p_{13}-p_{23}-1\}\ .
\end{align}

Using $(\ref{eq:sub1})$ and $(\ref{eq:sub2})$, one can easily show that each
number that appears in  $min\{\ldots\}$ of  $(\ref{eq:eta1})$ is greater or
equal to each number that appears in $max\{\ldots\}$ of (\ref{eq:eta2}), the
same for $(\ref{eq:chi1})$ and $(\ref{eq:chi2})$. Therefore,
$(\ref{eq:eta1})-(\ref{eq:chi2})$ define two non-empty intervals where one can
choose $\chi$ and $\eta$.  We can now write explicitly the coefficients
$\lambda(\varepsilon)$
 \begin{align*}
  \lambda(0,0,0)=&1-(p_1+p_2+p_3-p_{13}-p_{23})+\chi\ , \\
\lambda(1,0,0)=&p_1-p_{13}-\chi,\\ \lambda(0,1,0)=&p_2-p_{23}-\chi\ ,\\
\lambda(0,0,1)=&\eta + p_3 -p_{13} - p_{23},\\ \lambda(1,1,0)=&\chi\ ,\\
\lambda(1,0,1)=&p_{13}-\eta,\\ \lambda(0,1,1)=&p_{23}-\eta\ ,\\
\lambda(1,1,1)=&\eta\ .
\end{align*}

It follows immediately that $\lambda(\varepsilon)\geq 0 $ 
for all $\varepsilon\in\{0,1\}^3$, 
and that $\sum_{\varepsilon\in\{0,1\}^3}\lambda(\varepsilon)=1$.
 To conclude one just has to show, by writing it explicitly, that
$\sum_{\varepsilon\in\{0,1\}^3}\lambda(\varepsilon)u_\varepsilon=p$ and then
apply Proposition \ref{prop:polit}.$\square$\\

It follows that, for three observables, there exists a
classical representation for any state also in the case in which
there is only a pair of compatible observables and in the case of three
incompatible observables.

In fact, in the case of three incompatible observables only $p_1$, $p_2$ and
$p_3$ are given, thus one can add $p_{13}$ and $p_{23}$ that satisfy
(\ref{eq:sub1}) and (\ref{eq:sub2}) and then apply the same argument as in the
proof of Theorem \ref{teo:est3}. The same argument also applies to the
case  in which there is only a pair of compatible observables. Finally, if
property \textit{(K-S)} is assumed, a classical representation exists also for
three pairwise compatible observables.

We can conclude, therefore, that \textit{for three quantum
mechanical observables a classical probabilistic model which reproduce all
observable correlations always exists}.

We now discuss the implication of the results for the case of three
observables to the analysis of the case of four.
\begin{teo}\label{teo:est4}
 Let $\mathfrak{B}$ be a Boolean algebra freely generated by
$\mathcal{G}=\{A_1,A_2,A_3,A_4\}$, and  $\mathfrak{B}_{ij}$,
$\mathfrak{B}_{ijk}$, be the subalgebras generated respectively by
$\{A_i,A_j\}$ and $\{A_i,A_j,A_k\}$.

 Consider
$f:\mathfrak{B}_{13}\cup\mathfrak{B}_{23}
\cup\mathfrak{B}_{14}\cup\mathfrak{B}_{24}\longrightarrow[0,1]$
such that $f_{|_{\mathfrak{B}_{13}}}$, $f_{|_{\mathfrak{B}_{23}}}$,
$f_{|_{\mathfrak{B}_{14}}}$ and $f_{|_{\mathfrak{B}_{24}}}$ are normalized
measures on such subalgebras.\\ Then $f$ is extensible to a normalized measure
on the algebra $\mathfrak{B}$ if and only if there exist two partial
extensions  $f^{123}$ and $f^{124}$, of
$f_{|_{\mathfrak{B}_{13}\cup\mathfrak{B}_{23}}}$ and
$f_{|_{\mathfrak{B}_{14}\cup\mathfrak{B}_{24}}}$ on the subalgebras
$\mathfrak{B}_{123}$ and $\mathfrak{B}_{124}$,  such that
$f^{123}_{|_{\mathfrak{B}_{12}}}\equiv f^{124}_{|_{\mathfrak{B}_{12}}}$.
\end{teo}
\textbf{Proof} One implication is obvious since if a measure that extends $f$
exists, then the two partial extensions exist and they coincide on the
intersection.

For the converse, we note, as in Theorem \ref{teo:est3}, that we can consider
without loss of generality  ${X=\{A_1,A_2,A_3,A_4,A_1\cap A_3, A_2\cap A_3,
A_1\cap A_4, A_2\cap A_4\}}$ and $f:X\longrightarrow[0,1]$ and then apply
Proposition \ref{prop:polit}; therefore we construct the vector
$p=(p_1,p_2,p_3,p_4,p_{13},p_{23},p_{14},p_{24})$ and find the coefficients
$\lambda(\varepsilon)$.

First, we apply  Theorem \ref{teo:est3} to the subalgebras
$\mathfrak{B}_{123}$ and $\mathfrak{B}_{124}$ and to
$f_{|_{\mathfrak{B}_{13}\cup\mathfrak{B}_{23}}}$ and
$f_{|_{\mathfrak{B}_{14}\cup\mathfrak{B}_{24}}}$, obtaining two partial
extensions $f^{123}$ and $f^{124}$, that are normalized measures on the
subalgebras $\mathfrak{B}_{123}$ and $\mathfrak{B}_{124}$.

Now we consider the vector
\begin{eqnarray*}&p'=(p_1',p_2',p_3',p_{12}',p_{13}',p_{23}')
\text{ , } p_i'=f^{123}(A_i)\text{ , } i=1,2,3\\
 &p_{ij}'=f^{123}(A_i\cap A_j), 1\leq i < j\leq 3.
\end{eqnarray*}
and note that, by Proposition \ref{prop:polit}, there exist
$u_\varepsilon=(\varepsilon_1,\varepsilon_2,\varepsilon_3,\varepsilon_1
\varepsilon_2,\varepsilon_1\varepsilon_3,\varepsilon_2\varepsilon_3)$
and $\lambda'(\varepsilon)$ such that
\begin{equation*}
 p'=\sum_{\varepsilon\in\{0,1\}^3} \lambda'(\varepsilon)u_\varepsilon \text{
 ,}\qquad \sum_{\varepsilon\in\{0,1\}^3} \lambda'(\varepsilon)=1,\qquad
 \lambda'(\varepsilon)\geq 0.
\end{equation*}
The same argument applies to
\begin{align*}&p''=(p_1'',p_2'',p_3'',p_{12}'',p_{13}'',p_{23}''),\text{ , }  p_i''=f^{124}(A_i)\text{ , } i=1,2,4\\
& p_{12}''=f^{124}(A_1\cap A_2),\quad p_{i3}''=f^{124}(A_i\cap A_4), \quad i=1,2,
\end{align*}
and we obtain $\lambda''(\varepsilon)$ such that
\begin{eqnarray*}
p''=\sum_{\varepsilon\in\{0,1\}^3} \lambda''(\varepsilon)u_\varepsilon \text{
, } \sum_{\varepsilon\in\{0,1\}^3} \lambda''(\varepsilon)=1 \text{ , }
\lambda''(\varepsilon)\geq 0.
\end{eqnarray*}
Now we can define, for
$\varepsilon=(\varepsilon_1,\varepsilon_2,\varepsilon_3,\varepsilon_4)\in
\{0,1\}^4$,
\begin{equation}\label{eq:costrmis4}
 \lambda(\varepsilon)=\lambda(\varepsilon_1,\varepsilon_2,\varepsilon_3,\varepsilon_4)=\frac{\lambda'(\varepsilon_1,\varepsilon_2,\varepsilon_3)
\lambda''(\varepsilon_1,\varepsilon_2,\varepsilon_4)}{\lambda'(\varepsilon_1,\varepsilon_2,0)+\lambda'(\varepsilon_1,\varepsilon_2,1)}
\end{equation}
if the denominator is different from zero, and $\lambda(\varepsilon)=0$
otherwise (it is not difficult to show that the definition of
$\lambda(\varepsilon)$ is independent of choosing $\lambda'$ or $\lambda''$ in
the denominator)

It is obvious that $\lambda(\varepsilon)\geq 0$; their sum is given by
\begin{eqnarray*}
&\sum_{\varepsilon\in\{0,1\}^4}\lambda(\varepsilon)=\sum_{\varepsilon_1,\varepsilon_2}\sum_{\varepsilon_3,\varepsilon_4}\frac{\lambda'(\varepsilon_1,\varepsilon_2,\varepsilon_3)
\lambda''(\varepsilon_1,\varepsilon_2,\varepsilon_4)}{\lambda'(\varepsilon_1,\varepsilon_2,0)+\lambda'(\varepsilon_1,\varepsilon_2,1)}=\\
=&\sum_{\varepsilon_1,\varepsilon_2}\frac{[\lambda'(\varepsilon_1,\varepsilon_2,0)+\lambda'(\varepsilon_1,\varepsilon_2,1)][\lambda''(\varepsilon_1,\varepsilon_2,0)+\lambda''(\varepsilon_1,\varepsilon_2,1)]}{\lambda'(\varepsilon_1,\varepsilon_2,0)+\lambda'(\varepsilon_1,\varepsilon_2,1)}=\\
=&\sum_{\varepsilon_1,\varepsilon_2}\lambda'(\varepsilon_1,\varepsilon_2,0)+\lambda'(\varepsilon_1,\varepsilon_2,1)=
\sum_{\varepsilon\in\{0,1\}^3}\lambda'(\varepsilon)=1
\end{eqnarray*}
In the same way one verifies that
$\sum_{\varepsilon\in\{0,1\}^4}\lambda(\varepsilon)\varepsilon_i=p_i$, for
$i=1,2,3,4$, and that\\
$\sum_{\varepsilon\in\{0,1\}^4}\lambda(\varepsilon)\varepsilon_i\varepsilon_j=p_{ij}$
for $i=1,2$ e $j=3,4$. $\square$\\

It is interesting to notice that equation (\ref{eq:costrmis4}) has  the form
of a conditional probability $Pr(A,B,C)\equiv
Pr'(A|B)Pr''(C|B)Pr'(B)=\frac{Pr'(A,B)Pr''(C,B)}{Pr'(B)}$, where $B$ is an
event described by $(\varepsilon_1,\varepsilon_2)$, $A$ an event described by
$\varepsilon_3$ and $C$ an event described by $\varepsilon_4$. This allows in
fact for a generalization (see Theorem \ref{teo:catena} below).

Another interesting remark is that, since by Theorem \ref{teo:est3} two
partial extensions for three observables are always possible, Bell's
inequalities can be seen as compatibility (exactly in the sense of 
 states on a PBA) conditions for such partial extensions.

Moreover, the above is the only set of compatibility relations consistent with
property \textit{(K-S)} in which classical representability does not follow
from Theorem \ref{teo:catena} below. Theorem \ref{teo:est4} together with
Theorem \ref{teo:catena} provide, therefore, \textit{a complete analysis for
the case of four quantum observables}.\\

In general, the correlation polytope approach to the extension problem is
computationally intractable \cite{Pit89}, but the use of PBAs and associated states
 provides new criteria of classical
representability for PPTs. In the next section we
shall discuss some results that allow for a simplification of the computation
of extensibility conditions in many non-trivial cases. Another
extensibility criterion based on Horn and Tarski's notion of \textit{partial
measures} \cite{Tarski} is presented in Appendix \ref{sec:ht}.

\section{\label{sec:new}New extensibility criteria}
In this section we shall show how classical representability may arise
algebraically, i.e. independently of states, in many
non-trivial cases and we shall present some techniques that allow for a
simplification of the computation of conditions of classical representability.

The following result is given by a generalization of the proof of 
Theorem \ref{teo:est4}
\begin{teo}\label{teo:catena}
 Let $\mathfrak{B}$ be the Boolean algebra freely generated by
$\{A_1,\ldots,A_n\}$, and let $\mathfrak{B}_1$ and  $\mathfrak{B}_2$ be the
subalgebras generated respectively by $\{A_1,\ldots,A_k\}$ and
$\{A_i,\ldots,A_n\}$, with $1\leq i\leq k\leq n$. Let $\mu_1$ and $\mu_2$ be
two normalized measures on $\mathfrak{B}_1$ and $\mathfrak{B}_2$, such that
$\mu_1$ coincides with $\mu_2$ on $\mathfrak{B}_1\cap\mathfrak{B}_2$. Then a
measure $\mu$ which extends $\mu_1$ and $\mu_2$ on $\mathfrak{B}$ exist.
\end{teo}

\textbf{Proof} Using the  bijective correspondence between the atoms of the
subalgebra $\mathfrak{B}_1$ and the vectors
$\varepsilon'=(\varepsilon'_1,\ldots,\varepsilon'_k)\in\{0,1\}^k$, given by
${a_{\varepsilon'}=(-1)^{1-{\varepsilon'}_1}A_1\cap\ldots\cap(-1)^{1-\varepsilon'_k}A_k}$
(see Lemma \ref{lemma:atomlib} below), we define the function
${f_1:\{0,1\}^k\longrightarrow [0,1]}$ as
$f_1(\varepsilon')\equiv\mu_1(a_{\varepsilon'}) $ for all
$\varepsilon'\in\{0,1\}^k$. For all $A\in\mathfrak{B}_1$  
$\mu_1(A)=\sum_{\varepsilon'\in I_A} f_1(\varepsilon')$, where
$I_A\equiv\{\varepsilon'\in\{0,1\}^k | a_{\varepsilon'}\subset A\}$, therefore
$\mu_1(\textbf{1})=1$ implies
\begin{equation}\label{eq:f1sum}
 \sum_{\varepsilon'\in\{0,1\}^k}f_1(\varepsilon')=1
\end{equation}

We apply the same procedure to the measure $\mu_2$, defining
${f_2:\{0,1\}^{^{n-i+1}}\longrightarrow[0,1]}$ as
$f_2(\varepsilon'')\equiv\mu_2(a_{\varepsilon''})$, where $\varepsilon''\in
\{0,1\}^{n-i+1}$ and $a_{\varepsilon''}$ is an atom of
$\mathfrak{B}_2$. Similarly, $\mu_2(\mathbf{1})=1$ implies
\begin{equation}\label{eq:f2sum}
 \sum_{\varepsilon''\in\{0,1\}^{n-i+1}}f_2(\varepsilon'')=1.
\end{equation}

Moreover, $\mu_{1|_{\mathfrak{B}_1\cap\mathfrak{B}_2}}\equiv
\mu_{2|_{\mathfrak{B}_1\cap\mathfrak{B}_2}}$ implies
\begin{eqnarray}\label{eq:f12inter}
 \nonumber &\sum_{\varepsilon_1,\ldots,\varepsilon_{i-1}}f_1(\varepsilon_1,\ldots,\varepsilon_{i-1},\varepsilon_{i}\ldots,\varepsilon_k)=\\
 &=\sum_{\varepsilon_{k+1},\ldots,\varepsilon_{n}}f_2(\varepsilon_i,\ldots,\varepsilon_k,\varepsilon_{k+1}\ldots,\varepsilon_n),
\end{eqnarray}
where $\varepsilon_j\in\{0,1\}$, $j=1,\ldots,n$. In fact, since
$\mathfrak{B}_1\cap\mathfrak{B}_2$ is the subalgebra generated by\\
$\{A_i,A_{i+1},\ldots,A_k\}$, $(\ref{eq:f12inter})$ follows from the two
possible ways of writing every atom $a_{\eta}$ of
$\mathfrak{B}_1\cap\mathfrak{B}_2$, with $\eta\in\{0,1\}^{k-i+1}$, as a sum of
atoms of $\mathfrak{B}_1$ or of $\mathfrak{B}_2$.

We define $f$ on $\{0,1\}^n$ as follows
\begin{equation}\label{eq:probcondz}
f(\varepsilon_1,\ldots,\varepsilon_n)\equiv\frac{f_1(\varepsilon_1,\ldots,\varepsilon_i,\ldots,
\varepsilon_k)f_2(\varepsilon_i,\ldots,\varepsilon_k,\ldots,\varepsilon_n)}{\sum_{\eta_1,\ldots,\eta_{i-1}}f_1(\eta_1,
\ldots,\eta_{i-1},\varepsilon_i,\ldots,\varepsilon_k)},
\end{equation}
if the denominator is different from $0$, and $f(\varepsilon)=0$
otherwise. Equivalently, by $(\ref{eq:f12inter})$, the denominator can be
written with $f_2$, instead of $f_1$, summed over the last $(n-k)$ variables.

A function $\mu$ on $\mathfrak{B}$ is induced by $f$
\begin{equation}
 \mu(A)=\sum_{\varepsilon\in H_A} f(\varepsilon)
\end{equation}
for all $A\in\mathfrak{B}$, where  $H_A\equiv\{\varepsilon\in\{0,1\}^n |
a_\varepsilon\subset A\}$ and $a_\varepsilon$ is an atom of $\mathfrak{B}$
given by Lemma \ref{lemma:atomlib}.

It is obvious that $\mu$ is non-negative and additive on disjoint elements;
normalization ($\mu(\mathbf{1})=1$) follows easily from $(\ref{eq:f1sum})$ and
$(\ref{eq:f2sum})$.

To conclude, we need to show that $\mu$ coincide with $\mu_1$ on
$\mathfrak{B}_1$ and with $\mu_2$ on $\mathfrak{B}_2$. Given
$A\in\mathfrak{B}_1$, it can be written either as $A=\bigcup_{\varepsilon'\in
I_A} a_{\varepsilon'}$, where $\varepsilon'\in\{0,1\}^k$, $a_\varepsilon'$ is
an atom of $\mathfrak{B}_1$ and $I_A$ is defined as above, or else, in terms
of the atoms of $\mathfrak{B}$, as $A=\bigcup_{\varepsilon\in H_A}
a_{\varepsilon}$. By Lemma \ref{lemma:atomlib},
$H_A=\{(\varepsilon',\tilde{\varepsilon})\in \{0,1\}^n|\varepsilon'\in I_A,$
$\tilde{\varepsilon}\in\{0,1\}^{n-k}\}$ hence, by using also
$(\ref{eq:f12inter})$, we obtain
\begin{eqnarray*}
&\mu(A)=\sum_{\varepsilon\in H_A} f(\varepsilon)=\sum_{\varepsilon'\in
I_A}\sum_{\varepsilon_{k+1},\ldots,\varepsilon_{n}}
f(\varepsilon',\tilde{\varepsilon})= \\  
&=\sum_{\varepsilon'\in
I_A}\sum_{\varepsilon_{k+1},\ldots,\varepsilon_{n}}\frac{f_1(\varepsilon'_1,\ldots,
\varepsilon'_i,\ldots,\varepsilon'_k)f_2(\varepsilon'_i,\ldots,\varepsilon'_k,\tilde{\varepsilon}_{k+1},\ldots,\tilde{\varepsilon}_n)}{\sum_{\eta_1,\ldots,\eta_{i-1}}f_1
(\eta_1,\ldots,\eta_{i-1},\varepsilon'_i,\ldots,\varepsilon'_k)}=\\
&=\sum_{\varepsilon'\in
I_A}\sum_{\varepsilon_{k+1},\ldots,\varepsilon_{n}}\frac{f_1(\varepsilon'_1,\ldots,
\varepsilon'_i,\ldots,\varepsilon'_k)f_2(\varepsilon'_i,\ldots,\varepsilon'_k,\tilde{\varepsilon}_{k+1},\ldots,\tilde{\varepsilon}_n)}{\sum_{\eta_{k+1},\ldots,\eta_{n}}f_2
(\varepsilon'_i,\ldots,\varepsilon'_{k},\eta_{k+1},\ldots,\eta_n)}=\\
&=\sum_{\varepsilon'\in
I_A}f_1(\varepsilon'_1,\ldots,\varepsilon'_i,\ldots,\varepsilon'_k)=\mu_1(A).
\end{eqnarray*}
The proof for $\mu_2$ is analogous. $\square$\\

The above theorem provides an extensibility criterion based only on
the partial Boolean structures of observables, i.e. on 
compatibility relations. 

In many other non-trivial cases a classical extension of a
PPT is in fact obtainable ``for free'', i.e. from partial Boolean
structures, without constraints on the states. In the following we 
describe such cases in terms of compatibility relations, by introducing a
\textit{compatibility graph} representation.

Consider a Boolean algebra $\mathfrak{B}$ freely generated by a set of
observables $\mathcal{G}=\{A_1,\ldots,A_n\}$ with a function $f$ defined on a
subset $\bigcup_i\mathfrak{B}_i\subset\mathfrak{B}$, where each
$\mathfrak{B}_i$ is freely generated by a subset $\mathcal{G}_i\subset
\mathcal{G}$ and $f_{|_{\mathfrak{B}_i}}$ is a normalized measure on
$\mathfrak{B}_i$. We shall represent compatibility relations as follows:\\

\begin{itemize}
 \item each node represents a subalgebra $\mathfrak{B}_i$ on which $f$ defines
a normalized measure, and is  depicted as an ellipse where the generators
$\mathcal{G}_i\subset \mathcal{G}$, of the subalgebra are indicated;
\item when for two subalgebras, generated respectively by $\mathcal{G}'\subset
\mathcal{G}$ and $\mathcal{G}''\subset\mathcal{G}$,  $f$ defines a normalized
measure on the subalgebra generated by $\mathcal{G}'\cup\mathcal{G}''$, we
shall depict an edge connecting the two corresponding nodes.
\end{itemize}
An example is depicted in fig.1.

\begin{figure}[h]
\includegraphics[width=0.35\textwidth]{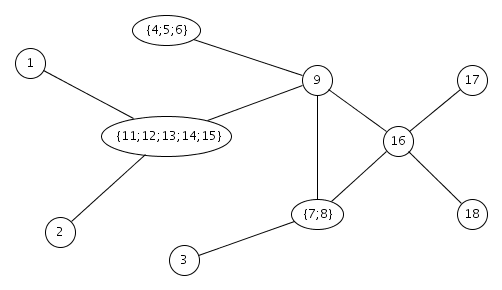}
\caption{ An example of a \textit{compatibility graph} representation}
\end{figure}

If  property \textit{(K-S)} holds, subgraphs consisting of pairwise 
connected nodes can be described by a single node. On one side, this allows 
for a representation using only single observable nodes,
on the other, it allows for a restriction to graphs without such 
completely connected subgraphs. An example is depicted in fig.2.

\begin{figure}[h]
        \includegraphics[width=0.22\textwidth]{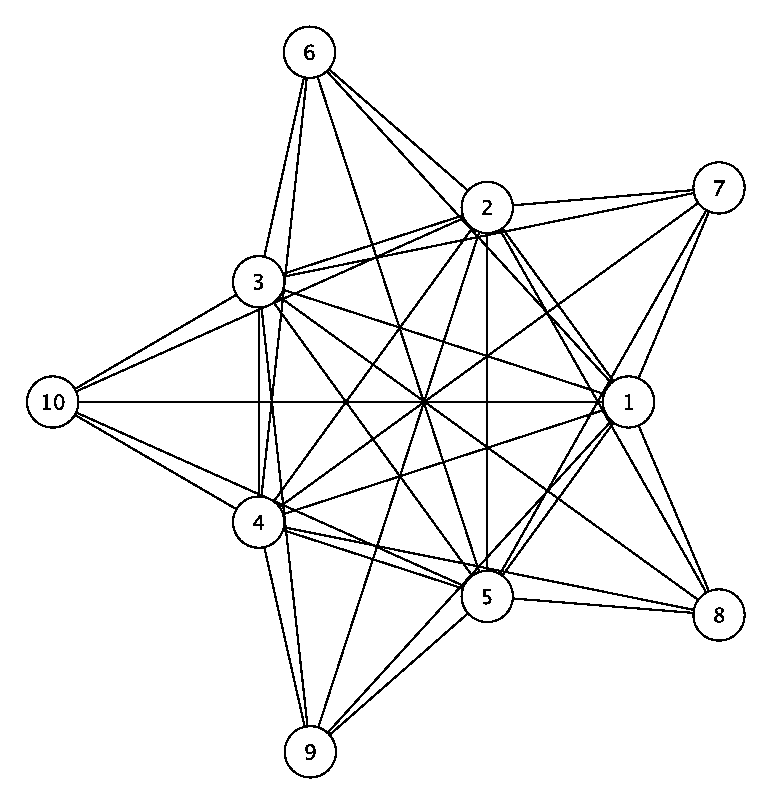}   
        \includegraphics[width=0.22\textwidth]{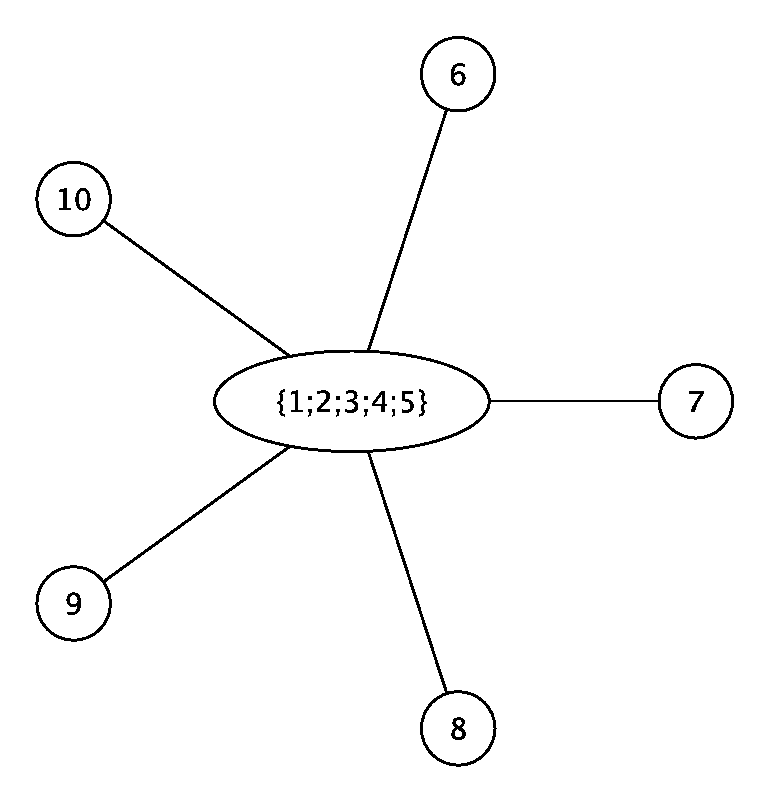}
 \caption{An example of two equivalent graph representations for a PPT where  property \textit{(K-S)} holds}
\end{figure}
A repeated application of Theorem \ref{teo:catena} gives the following
\begin{teo}\label{teo:tree}
Every PPT in which the compatibility graph is a tree graph,
i.e. a graph in which any two vertices are connected by exactly one path, is
extensible to a classical theory.
\end{teo}
As a consequence of Proposition \ref{prop:algdisg} Theorem \ref{teo:tree} 
applies also in the case of a collection of disconnected trees.\\

Even if quantum predictions for single-particle observables in 
entangled systems do not present 
\textit{tree-like} compatibility relations, 
the application of Theorem \ref{teo:tree} to
tree subgraphs substantially simplifies the analysis, as shown by the
following examples:

\begin{es}\label{es:bell} Starting from the result of Theorem \ref{teo:est4} and with the same notation, we discuss explicitly the conditions of extensibility of the  state $f$, namely the conditions guaranteeing the existence of two partial extension $f^{123}$ and $f^{124}$ that coincide on $\mathfrak{B}_{12}$. By Lemma \ref{lemma:misunic}, it is sufficient (see also proposition \ref{prop:polit}) that the two measure coincide on the element $A_1\cap A_2$. It is therefore sufficient to investigate possible attribution to $p_{12}$, by means of the Bell-Wigner polytope (see Pitowsky \cite{Pit89}), i.e. the correlation polytope associated with the subset  $X_s=\{A_1,A_2,A_s,A_1\cap A_s,A_2\cap A_s,A_1\cap A_2\}$, with $s=3,4$. One obtains\cite{Pit89} two systems of inequalities that can be written as
\begin{equation*}
 \alpha^{(3)}\leq p_{12}\leq \beta^{(3)},\qquad \alpha^{(4)}\leq p_{12}\leq
 \beta^{(4)} 
\end{equation*}
where $\alpha^{(s)}$ is the maximum of linear combinations $L^{(s)}$, of
$p_1,p_2,p_s,p_{1s},p_{2s}$, that appear in inequalities of the form
$L^{(s)}\leq p_{12}$, and $\beta^{(s)}$ is the minimum of of linear
combinations $L'^{(s)}$, of $p_1,p_2,p_s,p_{1s},p_{2s}$, that appear in
inequalities of the form $p_{12}\leq L'^{(s)}$. Therefore, a necessary and
sufficient condition for the extensibility of $f$ to a normalized measure on
$\mathfrak{B}$, i.e. for the classical representability of the corresponding
PPT, can be expressed as
\begin{equation}\label{eq:alfabeta}
 \max_{s=3,4} \alpha^{(s)}\leq \min_{s=3,4}\beta^{(s)}.
\end{equation}
By exchanging the role of pairs $1,2$ and $3,4$ one obtains an analogous
solution. It is less obvious, however, that (by Theorem \ref{teo:est4})
condition (\ref{eq:alfabeta}) for a consistent attribution of a value to the
correlation $p_{12}$ is satisfied if and only if the analogous condition for
the correlation $p_{34}$ is satisfied.

As a result, Bell inequalities can be identified as conditions for extensions
of states beyond their automatic extension given by algebraic structures.
Even if automatic extensions always exist, the violation of Bell inequalities 
shows the inconsistency of an absolute frequency interpretation.
 
\end{es}

\begin{es}\label{es:bellext}
Now consider a PPT as before but with observables
$A_5,\ldots,A_n$ added only ``on one side'', i.e. only compatible with $A_1,
A_2$.
More precisely, consider a Boolean
algebra $\mathfrak{B}$ freely generated by $\{A_1,\ldots,A_n\}$ and a function
$f:\bigcup_{i=1,2;j=3,\ldots,n}\mathfrak{B}_{ij}\longrightarrow[0,1]$, such
that $f_{|_{\mathfrak{B}_{ij}}}$ is a normalized measure on
$\mathfrak{B}_{ij}$ for $i=1,2$ $j=3,4,\ldots,n$.

Classical representability is still guaranteed, with the same notation as
before, by a condition analogous to (\ref{eq:alfabeta}), namely
\begin{equation}
  \max_{s=3,\ldots,n} \alpha^{(s)}\leq \min_{s=3,\ldots,n}\beta^{(s)},
\end{equation}
with a large simplification with respect to the general correlation 
polytope approach; in the present approach, the asymmetry 
of the problem allows in fact a discussion in terms of the 
``additional correlations'' for the ``side with less observables''.
\end{es}
A slightly more complicated example is the following:
\begin{es}\label{es:32} Let $\mathfrak{B}$ be a Boolean algebra freely generated
 by $\mathcal{G}=\{A_1,\ldots,A_6\}$ and $\mathfrak{B}_{ij}$ the subalgebra 
 generated by $\{A_i,A_j\}$, consider a function 
${f:\bigcup_{i=1,2,3;j=4,5,6}\mathfrak{B}_{ij}\longrightarrow[0,1]}$, such
that $f_{|_{\mathfrak{B}_{ij}}}$ is a normalized measure on
$\mathfrak{B}_{ij}$, with $i=1,2,3$ and $j=4,5,6$. By Theorem
\ref{teo:catena}, there exist three partial extensions $f^{123s}$ respectively
of $f_{|{\bigcup_{i=1}^3\mathfrak{B}_{is}}}$, $s=4,5,6$, to normalized
measures on $\mathfrak{B}_{123s}$, the subalgebra generated
$\{A_1,A_2,A_3,A_s\}$. The existence of an extension of $f$ to a normalized
measure on $\mathfrak{B}$ is equivalent to the condition that the three
partial extension $f^{123s}$ can be taken to coincide on the subalgebra
$\mathfrak{B}_{123}$. Such a condition can be investigated  in terms of the
values assumed by $\{A_1\cap A_2,A_1\cap A_3,A_2\cap A_3,A_1 \cap A_2\cap A_3\}$
and by means of the correlation polytopes associated to the sets
$X_s=\{A_1,A_2,A_3,A_s,A_1\cap A_s,A_2\cap A_s,A_3\cap A_s,A_1\cap A_2,A_1\cap
A_3,A_2\cap A_3,A_\cap A_2\cap A_3\}$, ${s=4,5,6}$. Each polytope is generated
by $16$ vertices in $\mathbb{R}^{11}$, and is described by $48$
inequalities, computed by means the double description method\cite{Motzk}
with \texttt{cdd} package \cite{Fukuda} (for the application of such methods
to QM predictions see Pitowsky and Svozil \cite{Pit-Svo}). 
However, out of these $48$ inequalities only $32$ are relevant
for our discussion and they can be written as: 
\begin{eqnarray*}
\text{Type 1 } \alpha^{(s)}_{1,ij}\leq p_{ij}\leq
\beta^{(s)}_{1,ij}\text{ }, \alpha^{(s)}_{1}\leq p_{123}\leq
\beta^{(s)}_1\\ 
\{ij\}=\{12\}, \{13\}, \{23\}\ ,\\ 
\text{Type 2 } \qquad \text{ } \qquad \text{ } \quad \alpha^{(s)}_{2,ij}\leq p_{ij}-p_{123}\leq \beta^{(s)}_{2,ij}\\
\{ij\}=\{12\}, \{13\}, \{23\}\ ,\\ 
\text{Type 3 } \qquad \alpha^{(s)}_{3,ijk}\leq p_{ij}+p_{jk}-p_{123}\leq \beta^{(s)}_{3,ijk}\\
\{ij\}=\{12\}, \{13\}, \{23\}, j\neq k\neq i\ ,\\ 
\text{Type 4 }
\alpha^{(s)}_4\leq p_{12}+p_{13}+p_{23}-p_{123}\leq \beta^{(s)}_4
\text{ ,}
\end{eqnarray*}
where $\alpha^{(s)}_{1,ij}$ is the maximum of of linear combinations
$L^{(s)}$, of $p_1,p_2,p_3,p_{s},p_{1s},p_{2s},p_{3s}$, that appear in
inequalities of the form $L^{(s)}\leq p_{ij}$, and $\beta^{(s)}$ is the
minimum of of linear combinations $L'^{(s)}$, of
$p_1,p_2,p_3,p_s,p_{1s},p_{2s},p_{3s}$, that appear in inequalities of the
form ${p_{ij}\leq L'^{(s)}}$, and so on for each type of inequalities. A
necessary and sufficient condition for the extensibility of $f$ is therefore
the existence of a solution for the following system of $11$ linear
inequalities in the variables $p_{12},p_{13},p_{23},p_{123}$,
\begin{align}\label{eq:t1}
 &\max_s\alpha^{(s)}_{1,ij}\leq p_{ij}\leq \min_s\beta^{(s)}_{1,ij},\\  
\nonumber&\{ij\}=\{12\}, \{13\}, \{23\};\\ &\max_s\alpha^{(s)}_{1}\leq p_{123}\leq
\min_s\beta^{(s)}_1;\\ &\max_s\alpha^{(s)}_{2,ij}\leq p_{ij}-p_{123}\leq
\min_s\beta^{(s)}_{2,ij}, \\
\nonumber &\{ij\}=\{12\}, \{13\}, \{23\};\\
&\max_s\alpha^{(s)}_{3,ijk}\leq p_{ij}+p_{jk}-p_{123}\leq
\min_s\beta^{(s)}_{3,ijk},\\
\nonumber &\{ij\}=\{12\}, \{13\}, \{23\}, j\neq k\neq i;\\
\label{eq:t4} &\max_s\alpha^{(s)}_4\leq p_{12}+p_{13}+p_{23}-p_{123}\leq \min_s\beta^{(s)}_4. &
\end{align}
As in example \ref{es:bell}, the discussion with $1,2,3$ substituted with
$4,5,6$ is analogous and the corresponding conditions are equivalent.

Moreover, as in example \ref{es:bellext}, the introduction of additional
observables only ``on one side'' modifies inequalities
(\ref{eq:t1})-(\ref{eq:t4}) only by extending the sets on which maximum and
minimum are taken.
\end{es}

In general, our method consists in exploiting the ``algebraic''  extensions
given by Theorem \ref{teo:tree}:  conditions of classical
representability only arise as consistency  (i.e. coincidence on
intersections) conditions for putting together partial 
extensions associated to tree subgraphs, giving rise to a description
of the initial compatibility graph as a tree graph on such extended nodes.
As pointed out in example \ref{es:bellext}, different strategies are possible, 
keeping the values of the given set of correlations fixed throughout all partial
extensions.\\

\section{Conclusions} We have presented an algebraic approach to the
extensions of partial probability theories, i.e. partial classical
structures, applicable in particular to those arising in QM, that contains in
the same logical framework both Bell-type and Kochen-Specker-type approaches.

The above analysis applies in particular to the problem of simulability of quantum
algorithms by means of classical (probabilistic) algorithms. It is well known
that a partial probability theory can be simulated  by a classical theory where
additional variables, representing observables in different contexts, are
introduced. Given a PPT, associated to a finite set of 
quantum measurements, it is less obvious whether and how much it can 
be extended; in other words whether the number of its contexts, i.e., 
the number of additional classical variables can be reduced.

The analysis of partial extensions of a PPT provides, therefore,
an intrinsic measure of its \textit{non-classicality}. From an
information-theoretic viewpoint, the above mentioned additional variables may
be interpreted as \textit{additional information} carried by a quantum system
with respect to its classical counterpart.

We stress that because of the well known computational intractability of the
correlation polytope approach \cite{Pit89} other approaches should be
investigated. 
Beyond the application of Theorem \ref{teo:catena} in examples 
\ref{es:bell}, \ref{es:bellext} and \ref{es:32}, in appendix C
we point out two additional methods: the first one is
to investigate the partial order $\leq$ of definition \ref{def:seqord} (see
Appendix \ref{sec:ht}) since it enters in the definition of partial measure and
provides conditions of extendibility; the second one is to investigate
properties of interior and exterior measures since they play a fundamental
role in Theorem \ref{teo:feg}.

We briefly outline some implications of our approach for the interpretations
of QM. It has been shown in example \ref{es:bell} that condition
(\ref{eq:alfabeta}) is equivalent to the complete (in the sense of Pitowsky\cite{Pit89})
set of CHSH inequalities; in other words that classical representability 
is equivalent to the possibility of an attribution of a value to the 
\textit{non-observable correlation} $p_{12}$ which is consistent with the observable
correlations. This example points out explicitly the role of 
predictions for non-observable correlations in every attempt 
to a classical interpretation of quantum
mechanical predictions. This
allows for an interpretation of the violation of Bell inequalities as
a negative answer to the question: given two incompatible
observables  is it possible to assign a value to their correlation which is
\textit{consistent with measurable correlations}?

It is interesting to notice that this consistency criterion is implicit
in the argument by Einstein Podolsky and Rosen. In fact, in their famous
paper \cite{EPR} they discussed the possibility of extending 
quantum mechanical predictions to include correlations between incompatible
observables with the requirement that such added (non-observable) correlations
are consistent with the observable correlations; in their case, a value is
attributed to the momentum of a particle on the basis of its {(perfect
anti-) correlation} with the momentum of the other particle, giving rise to a
joint attribution of definite values to position and momentum of a single
particle. We may, therefore, name such consistency conditions as \textit{the
EPR criterion for extensions of QM predictions}.

\appendix
\section{\label{sec:ba}Boolean Algebras}
There exists a vast literature on Boolean algebras that explores deep aspects
of the subject and important connections with several branches of mathematics;
we briefly recall in this section the elementary results needed
for our discussion, involving in particular only finite sets.
For more details see
Sikorski \cite{Sikorski} and Givant-Halmos \cite{G-H}.\\
 
A Boolean algebra is a non-empty set $\mathfrak{B}$ in which 
two binary operations $\cap$ and $\cup$ are defined ,  
called respectively \textit{meet} and
\textit{join}, and one unary operation $^c$ called \textit{complement},
satisfying  certain axioms.

We shall denote by $\emptyset$ the \textit{zero} element, 
 $\emptyset = A\cap A^c$ for all $A$,  and by $\textbf{1}$
the \textit{unit}, $ \textbf{1}= A\cup A^c$ for all $A$.

Every finite Boolean algebra is isomorphic to the Boolean algebra
$\mathcal{P}(X)$ (the subsets of $X$ with intersection, union and complement) 
for a certain finite set $X$. 

 Given a Boolean algebra $\mathfrak{B}$, a function
$m:\mathfrak{B}\longrightarrow\mathbb{R}$ is called a  \textit{measure} on
$\mathfrak{B}$ if it satisfies:
\begin{itemize}
 \item[$(a)$] $0\leq m(A)\leq \infty$ for all $A\in\mathfrak{B}$, and there exists
$A_0\in\mathfrak{B}$ such that  $m(A_0)<\infty$;
\item[$(b)$] $m(A\cup B)=m(A) + m(B)$, if $A$ and $B$ are two disjoint
elements of $\mathfrak{B}$.
\end{itemize}
A measure that satisfy $m(\textbf{1})=1$ is called a \textit{normalized
measure}, and a normalized measure such that $m(A)\in\{0,1\}$ for every
$A\in\mathfrak{B}$ is called a \textit{two-valued measure} or a
\textit{multiplicative measure}, since its properties also imply that $m(A\cap
B)=m(A)m(B)$, for all $A,B\in\mathfrak{B}$.

 An element $a\neq\emptyset$ of a Boolean algebra $\mathfrak{B}$ is called an
\textit{atom} if, for all  $A\in\mathfrak{B}$ the inclusion $A\subset a$
implies $A=\emptyset$ or $A=a$. A Boolean algebra $\mathfrak{B}$ is called
\textit{atomic} if for every element $A$ in $\mathfrak{B}$ exists an atom $a$
such that $a\subset A$.
\begin{lemma}\label{lemma:atom}
Every finite Boolean algebra $\mathfrak{B}$ is atomic and, if it has $N$
elements, it has exactly $n$ atoms $\{a_1,\ldots,a_n\}$ such that
$N=2^n$. Moreover, every element $A\in\mathfrak{B}$ can be written uniquely as
$A=\bigcup_{i\in I}a_i$, where $I\subset\{1,\ldots,n\}$.
\end{lemma}
\begin{lemma}\label{lemma:mismolt}
 Let $\mathfrak{B}$ be a finite Boolean algebra with $n$ atoms
$\{a_1,\ldots,a_n\}$; then for every atom $a_i$ there exists a  multiplicative
measure $\delta_{a_i}$ which is $1$ on $a_i$ and $0$ on all other
atoms. Moreover multiplicative measures on $\mathfrak{B}$ are all and only
those that are defined in this way.
\end{lemma}
\textbf{Proof} Given the atom $a_i$, define the function
$\delta_{a_i}:\mathfrak{B}\longrightarrow\{0,1\}$ as follows:
$\delta_{a_i}(A)=1$ if $a_i\subset A$, and zero otherwise. It follows
immediately that $\delta_{a_i}(A)\in\{0,1\}$, while for the property
$\delta_{a_i}(A\cup B)=\delta_{a_i}(A)+\delta_{a_i}(B)$, if $A\cap
B=\emptyset$, it is sufficient to check that it holds in the three possible
cases: $a_i\subset A$ , $a_i\subset B$ and $a_i\subset (A\cup B)^c$
($a_i\subset A\cap B$ is excluded since $A\cap B=\emptyset$). Therefore,
$\delta_{a_i}$ is a two-valued measure.
The result follows from a check that there are no other multiplicative
measures. $\square$\\

Given a Boolean algebra $\mathfrak{B}$, a subset
$\mathcal{G}\subset\mathfrak{B}$ is said to be a set of \textit{generators}
for $\mathfrak{B}$ if, for all $B\in\mathfrak{B}$, $B$ can be represented in
the form
\begin{equation}
 B=(A_{1,1}\cap\ldots\cap A_{1,r_1})\cup\ldots\cup(A_{s,1}\cap\ldots\cap A_{s,r_s}),
\end{equation}
  where for all $m,n$ either $A_{m,n}\in\mathcal{G}$ or
  $A_{m,n}^c\in\mathcal{G}$.

A set $\mathcal{G}$ of generators of a Boolean algebra $\mathfrak{B}$ is said
to be \textit{free} if every mapping from $\mathcal{G}$ to an arbitrary
Boolean algebra $\mathfrak{B}'$ can be extended to a $\mathfrak{B}'$-valued
homomorphism on $\mathfrak{B}$. Moreover, a Boolean algebra is said to be
\textit{freely generated} or simply \textit{free} if it contains a set of free
generators.

\begin{lemma}\label{lemma:atomlib}
 Given a free Boolean algebra $\mathfrak{B}$ with $n$ free generators
$\mathcal{G}=\{A_1,\ldots,A_n\}$, it contains  $2^n$ atoms $a_\varepsilon$
which are given by the possible intersections $\bigcap_{i=1}^n
(-1)^{1-\varepsilon_i} A_i$, where $-A_i\equiv A_i^c$ and
$\varepsilon=(\varepsilon_1,\ldots,\varepsilon_n)\in\{0,1\}^n$.

Moreover, given a subalgebra $\mathfrak{B}_0$ generated by
$\{A_1,\ldots,A_k\}$, with $1<k<n$, every atom of $\mathfrak{B}_0$, which can
be written as $b_{\varepsilon'}=(-1)^{1-\varepsilon'_1}A_1\cap\ldots\cap
(-1)^{1-\varepsilon'_k}A_k$, with
$\varepsilon'=(\varepsilon'_1,\ldots,\varepsilon'_k)\in\{0,1\}^k$, can be
written in terms of atoms $a_{(\varepsilon',\tilde{\varepsilon})}$ of
$\mathfrak{B}$, with $\tilde{\varepsilon}\in \{0,1\}^{n-k}$ and
$(\varepsilon',\tilde{\varepsilon})\in \{0,1\}^n$, as
\begin{equation}\label{eq:atomsubalgebra}
b_{\varepsilon'}=\bigcup_{\tilde{\varepsilon}\in \{0,1\}^{n-k}}
a_{(\varepsilon',\tilde{\varepsilon})}\qquad.
\end{equation}

\end{lemma}
\textbf{Proof} First we notice that a multiplicative measure on $\mathfrak{B}$
is a homomorphism between $\mathfrak{B}$ and the set $\{0,1\}$ with Boolean
operations defined as $x\cap y=xy$, $x\cup y=x+y-xy$ and $x^c=1-x$ for all
$x,y\in\{0,1\}$. Therefore, it follows from definition of free Boolean algebra
that each map $f:\mathcal{G}\longrightarrow\{0,1\}$ can be extended, in a
unique way since $\mathcal{G}$ is a set of generators, to a multiplicative
measure.

All possible $\{0,1\}-valued$ maps on $\mathcal{G}$, are labeled by
$\varepsilon=(\varepsilon_1,\ldots,\varepsilon_n)\in\{0,1\}^n$,
i.e. $f_\varepsilon:\mathcal{G}\longrightarrow\{0,1\}$, where such a
correspondence is given by $f_\varepsilon(A_i)=\varepsilon_i$ . Since each map
can be extended in a unique way to a multiplicative measure on $\mathfrak{B}$,
there is a one-to-one correspondence, by Lemma \ref{lemma:mismolt}, between
$\varepsilon\in\{0,1\}$ and the atoms of $\mathfrak{B}$. Let $m_\varepsilon$
be the multiplicative measure that extends $f_\varepsilon$; from condition
$m_\varepsilon(A_i)=\varepsilon_i$ we obtain
$m_\varepsilon((-1)^{1-\varepsilon_i}A_i)=1$, which implies
$m_\varepsilon(\bigcap_{i=1}^n (-1)^{1-\varepsilon_i}A_i)=1$. It is then
enough to verify
that for every $\varepsilon$ the element $\bigcap_{i=1}^n
(-1)^{1-\varepsilon_i}A_i$ is an atom and that they are all distinct, 
since they are in the right number for a bijection 
with multiplicative measures.

It is easy to show that $ \bigcap_{i=1}^n (-1)^{1-\varepsilon_i}A_i\cap
\bigcap_{i=1}^n (-1)^{1-\varepsilon'_i}A_i=\emptyset$ if
$\varepsilon\neq\varepsilon'$, in fact, it implies
$\varepsilon_j\neq\varepsilon'_j$ for at least one $j$, therefore in the above
product $A_j\cap A_j^c$ must appear.

Now consider a generic $a_{\varepsilon'}$, we show that $\forall B\in
\mathfrak{B}$ either $ a_{\varepsilon'}\subset B$, or ${B\cap
a_{\varepsilon'}=\emptyset}$. Since $\mathfrak{B}$ is a free algebra, for all
$B$ we can write $B=\bigcup_{j=1}^k\bigcap_{i \in
I_j}(-1)^{1-\varepsilon_i^j}A_i$, where $k\in\mathbb{N}$,
$\varepsilon_i^j\in\{0,1\}$ and $I_j\subset \{1,\ldots,n\}$. Now, for fixed $j$,
either $\varepsilon'_i=\varepsilon_i^j$ for all $i \in I_j$, and then
$a_{\varepsilon'}\subset( \bigcap_{i \in I_j}(-1)^{1-\varepsilon_i^j}A_i)$, or
there exists $i$ such that $\varepsilon'_i\neq\varepsilon_i^j$ and then
$a_{\varepsilon'}\cap (\bigcap_{i \in
I_j}(-1)^{1-\varepsilon_i^j}A_i)=\emptyset$. By repeating this argument for
all $j$ we obtain that either $a_{\varepsilon'}$ is contained in at least one
of the $(\bigcap_{i \in I_j}(-1)^{1-\varepsilon_i^j}A_i)$, and then in $B$, or
it has empty intersection with all of them and therefore $a_{\varepsilon'}\cap
B=\emptyset$. Therefore $a_{\varepsilon'}$ is an atom.

For the second part, it is sufficient to write explicitly equation
(\ref{eq:atomsubalgebra}) in terms of generators of $\mathfrak{B}$ and use
iteratively the equation $A=(A\cap B)\cup (A\cap B^c)$. $\square$

\section{\label{sec:cp}Correlation polytopes}

In this appendix we shall derive extension criteria which follow from the
translation of Pitowsky's correlation polytopes results into the Boolean
framework.

First, we  briefly outline Pitowsky's result (for pair correlations): given
 $n$ atomic propositions $a_1,\ldots,a_n$ and positive numbers
 $p_1,\ldots,p_n,\ldots p_{ij}\ldots$, $\{ij\}\in S\subset \{ \{ij\} | 1\leq
 i<j \leq n \}$ associated to the propositions $a_1,\ldots,a_n$ and some
 logical conjunction of them $(a_i \wedge a_j)$, $\{ij\}\in S$, the numbers
 $p_i, p_{ij}$ are interpretable in terms of classical probabilities
 (i.e. there exists a probability space $(X,\Sigma, \mu)$ and $n$ events
 $A_1,\ldots,A_n$ such that $p_i=\mu(A_i)$, $p_{ij}=\mu(A_i\cap A_j)$) if and
 only if the vector $(p_1,\ldots,p_n,\ldots,p_{ij},\ldots)$ is a convex
 combination of the vectors ${u_\varepsilon=(\varepsilon_1,\ldots,
 \varepsilon_n,\ldots,\varepsilon_i\varepsilon_j,\ldots)}$,
 $\varepsilon=(\varepsilon_1,\ldots,\varepsilon_n)\in\{0,1\}^n$.

The problem can be expressed in terms of free Boolean algebras with the
identification (see Givant and Halmos \cite{G-H}) of  atomic propositions with
free generators, logical operations with Boolean operations, truth assignments
with two-valued measures, and probability assignment with normalized measures.
A basic fact is the following: 

\begin{lemma}\label{lemma:normmeas}
 Let $\mathfrak{B}$ be a finite Boolean algebra with $n$ atoms
 $\{a_1,\ldots,a_n\}$; then a function $m:\mathfrak{B}\longrightarrow [0,1]$ is
 a normalized measure $\Longleftrightarrow$ $m=\sum_{i=1}^n \lambda_i
 \delta_{a_i}$, where $\delta_{a_i}$ is the multiplicative measure which is
 $1$ on $a_i$, and $\lambda_i\geq 0$
 and $\sum_{i=1}^n\lambda_i=1$.
\end{lemma}

\textbf{Proof}.  Given $A\in\mathfrak{B}$,    it can be written  as a 
disjoint union $A=\bigcup_{i\in I}a_i$, $I\subset\{1,\ldots,n\}$. Then
\begin{equation}
 m(A)=\sum_{i\in I}m(a_i)=\sum_{i\in I} m(a_i)\delta_{a_i}(a_i)=\sum_{i=1}^n
 m(a_i)\delta_{a_i}(A).
\end{equation}
Moreover $0\leq m(a_i)\leq 1$, since $m$ is a normalized measure, and
$\sum_{i=1}^n m(a_i)=m(\bigcup_{i=1}^n a_i)=m(\textbf{1})=1$.

The converse is obvious.$\square$\\

From this lemma we can obtain a criterion of extensibility to normalized
measures for functions defined over a subset of a finite Boolean algebra. In
fact, if we consider a finite Boolean algebra $\mathfrak{B}$ with $k$ atoms
$\{a_1,\ldots,a_k\}$ and a subset $X\subset \mathfrak{B}$, then a function
$f:X\longrightarrow[0,1]$ can be extended to normalized measure $\mu$ on
$\mathfrak{B}$ if and only if there are $k$ numbers
$\lambda_1,\ldots,\lambda_k$, with $\lambda_i\geq 0$ and $\sum_{i=0}^k
\lambda_i =1$, such that
\begin{equation}
\label{eq:est} f=\sum_{i=1}^k \lambda_i {\delta_{a_i}}_{|_X}
 \end{equation}
and $\mu$ is given by:
\begin{equation}
 \mu=\sum_{i=1}^k \lambda_i \delta_{a_i}
\end{equation}
Therefore assignments of values in $[0,1]$ to elements of a subset of a
Boolean algebra have a probabilistic interpretation if and only if such values
are given by a convex combination of two-valued measures. This construction is
closely related to the notion of \textit{correlation polytopes}, as shown
by the following
\begin{prop}\label{prop:polit}
  Let $\mathfrak{B}$ be a Boolean algebra freely generated by
$\mathcal{G}=\{A_1,\ldots,A_n\}$ and consider $X\subset \mathfrak{B}$ with
$X=\{A_1,\ldots,A_n,\ldots,A_i\cap A_j,\ldots,A_i\cap A_j\cap A_k,\ldots\}$,
i.e., $X=\mathcal{G}\cup S_2\cup\ldots S_m$, $m\leq n$, where elements of $S_l$
are the intersections of $l$ distinct generators, but not necessarily all of
those possible, i.e. $|S_l|\leq \binom{n}{l}$. Now consider
$f:X\longrightarrow[0,1]$ and define the vector
\begin{equation*} 
p=(p_1,\ldots,p_n,\ldots p_{ij},\ldots,p_{i_1\ldots i_m},\ldots)\in
\mathbb{R}^{|X|}
\end{equation*}
 which has as components the values assumed by $f$ on $X$, namely
\begin{eqnarray*} p_i=f(A_i),&\quad &i=1,\ldots,n,\\
 p_{ij}=f(A_i\cap A_j),&\quad &A_i\cap A_j\in S_2\\ &\vdots \quad &\vdots\\
p_{i_1\ldots i_m}=f(A_{i_1}\cap\ldots \cap A_{i_m}),&\quad
&A_{i_1}\cap\ldots\cap A_{i_m}\in S_m.
 \end{eqnarray*}
  For every $\varepsilon \in \{0,1\}^n$ define the vector $u_{\varepsilon}\in
  \{0,1\}^{|X|}$ given by
\begin{displaymath}
 u_\varepsilon = (\varepsilon_1,\ldots,\varepsilon_n,\ldots,\varepsilon_i
 \varepsilon_j,\ldots,
 \varepsilon_{i_1}\varepsilon_{i_2}\ldots\varepsilon_{i_m},\ldots)
\end{displaymath}
i.e. for every component $p_{i_1\ldots i_k}$ of $p$ there is a corresponding
component of $u_\varepsilon$ given by
$\varepsilon_{i_1}\ldots\varepsilon_{i_k}$.

Then $f$ can be extended to a normalized measure on $\mathfrak{B}$ if and only
if there are $2^n$ numbers $\lambda(\varepsilon)$, $\varepsilon\in\{0,1\}^n$,
such that:
\begin{displaymath}
  p=\sum_{\varepsilon\in\{0,1\}^n} \lambda(\varepsilon) u_\varepsilon
\end{displaymath}
\begin{displaymath}
 \text{ with } \lambda(\varepsilon) \geq 0 \text{ } \forall \varepsilon \in
 \{0,1\}^n,\text{ and }\sum_{\varepsilon \in \{0,1\}^n}
 \lambda(\varepsilon)=1
\end{displaymath}
\end{prop}
\textbf{Proof} Given a free Boolean algebra with $n$ free generators,  there
are (see Lemma \ref{lemma:atomlib}) $2^n$ atoms in a one-to-one
correspondence with $\varepsilon\in\{0,1\}^n$. The extensibility condition
(\ref{eq:est}) can therefore be written, with $a_\varepsilon$ defined as in
Lemma \ref{lemma:atomlib},
\begin{equation}
 f=\sum_{i=1}^{2^n}\lambda_i {\delta_{a_i}}_{|_X}=\sum_{\varepsilon \in
 \{0,1\}^n}\lambda(\varepsilon){\delta_{a_\varepsilon}}_{|_X}
\end{equation}
Now it suffices to verify such a condition for every element of $X$, which is
equivalent to verify that the vector $p$ is a convex combination of the
vectors $u_\varepsilon$. In fact, $\delta_{a_\varepsilon}(A_i)=\varepsilon_i$
(see the proof of Lemma \ref{lemma:atomlib}) and
${\delta_{a_\varepsilon}(A_{i_1}\cap\ldots\cap
A_{i_k})=\varepsilon_{i_1}\ldots\varepsilon_{i_k}}$, therefore:
\begin{widetext}
\begin{align*}
p_i=\sum_{\varepsilon \in \{0,1\}^n}\lambda(\varepsilon)\varepsilon_i \quad
&\Longleftrightarrow \quad f(A_i)=\sum_{\varepsilon \in
\{0,1\}^n}\lambda(\varepsilon)\delta_{a_\varepsilon}(A_i)\quad i=1,\ldots,n,\\
p_{ij}=\sum_{\varepsilon \in
\{0,1\}^n}\lambda(\varepsilon)\varepsilon_i\varepsilon_j \quad
&\Longleftrightarrow \quad f(A_i\cap A_j)=\sum_{\varepsilon \in
\{0,1\}^n}\lambda(\varepsilon)\delta_{a_\varepsilon}(A_i\cap A_j), \quad
A_i\cap A_j \in S_2\\ &\vdots \qquad \vdots\\ p_{i_1\ldots
i_m}=\sum_{\varepsilon \in
\{0,1\}^n}\lambda(\varepsilon)\varepsilon_{i_1}\ldots\varepsilon_{i_m}&\Longleftrightarrow
f(A_{i_1}\cap\ldots\cap A_{i_m})=\sum_{\varepsilon \in
\{0,1\}^n}\lambda(\varepsilon)\delta_{a_\varepsilon}(A_{i_1}\cap\ldots\cap
A_{i_m}),\\  &\qquad\qquad\qquad\qquad\qquad\qquad A_{i_1}\cap\ldots\cap
A_{i_m} \in S_m\quad .\quad \square
\end{align*}
\end{widetext}

We  denote the correlation polytope associated with a
subset $X\subset\mathfrak{B}$ of the above mentioned form as
$C(\mathcal{G},S_2,\ldots,S_m)$.

A helpful fact about the characterization of measures is the following

\begin{lemma}\label{lemma:misunic}
 Let $\mathfrak{B}$ be a Boolean algebra freely generated by
$\{A_1,\ldots,A_n\}$, let $\mu$ be a measure on  $\mathfrak{B}$ and let
$X\subset \mathfrak{B}$ the set of all possible intersections between the free
generators of $\mathfrak{B}$, i.e. $X=\{A_1,\ldots,A_n,\ldots,A_i\cap
A_j,\ldots, A_{i_1}\cap\ldots\cap A_{i_k},\ldots,A_1\cap\ldots\cap
A_n\}$. Then the measure $\mu$ is uniquely defined by the values it assumes on
the set $X$.
\end{lemma}
\textbf{Proof} Since $\mathfrak{B}$ is a finite algebra, it is
atomic. Therefore every element $B\in\mathfrak{B}$ can be written in a unique
way as a disjoint union of atoms, $B=\bigcup_{\varepsilon\in H_B}
a_\varepsilon$, where
$\varepsilon=(\varepsilon_1,\ldots,\varepsilon_n)\in\{0,1\}^n$,
$a_\varepsilon=\bigcap_i(-1)^{1-\varepsilon_i}A_i$ is the atom given by the
bijection of Lemma \ref{lemma:atomlib} and
$H_B\equiv\{\varepsilon\in\{0,1\}^n$ $|$ $a_\varepsilon\subset B$ $\}$. It follows
that $\mu(B)=\sum_{\varepsilon\in H_B} \mu(a_\varepsilon)$ since atoms are
disjoint, therefore the measure is completely defined by the values assumed on
atoms.

Now, it suffices to prove that the measure of every atom is uniquely
determined by the values assumed by $\mu$ on the elements of $X$. But this
follows from the fact that every atom can be written in the form
$a_\varepsilon=\bigcap_i(-1)^{1-\varepsilon_i}A_i$ and the property of
measures
\begin{equation}\label{eq:smdisg}
\mu(A)=\mu(A\cap B)+\mu(A\cap B^c) \text{ for all }A,B\in\mathfrak{B}.
\end{equation}
 In fact, consider an atom
$a_\varepsilon=(-1)^{1-\varepsilon_1}A_1\cap\ldots\cap(-1)^{1-\varepsilon_n}A_n$;
if  $\varepsilon_i=1$ for all $i$, then $a_\varepsilon\in X$ so there is
nothing to show. Assume, therefore, that there is $i_0$ such that
$\varepsilon_{i_0}=0$; thus, for (\ref{eq:smdisg}), $\mu(a_\varepsilon)$ can
be written in terms of elements of $X$ as
$\mu(a_\varepsilon)=\mu(\bigcap_{i\neq i_0} A_i)-\mu(\bigcap_i A_i)$.

If there is more than one component of $\varepsilon$ equal to $1$, this
process can be iterated until $\mu(a_\varepsilon)$ is written in terms of
elements of $X$.$\square$\\

The above general results provide a criterion for the extensibility of
functions defined over a subset of a free Boolean algebra to normalized
measures on the entire algebra, and therefore a criterion of classical
representability for PPTs.
A simple application is given by the following result:

\begin{prop}\label{prop:algdisg}
 Let $\mathfrak{B}$ be the Boolean algebra freely generated by
$\{A_1,\ldots,A_n\}$, let $\mathfrak{B}_1$ and  $\mathfrak{B}_2$ be the
subalgebras generated respectively by  $\{A_1,\ldots,A_k\}$ and
$\{A_{k+1},\ldots,A_n\}$ with $1\leq k< n$, and let $\mu_1$ and $\mu_2$ be two
normalized measure respectively on $\mathfrak{B}_1$ and $\mathfrak{B}_2$. Then
there exists a normalized measure $\mu$ on $\mathfrak{B}$  such that
$\mu_{|_{\mathfrak{B}_1}}\equiv \mu_1$ and $\mu_{|_{\mathfrak{B}_2}}\equiv
\mu_2$.
\end{prop}
\textbf{Proof} Every atom $a_\varepsilon$, $\varepsilon\in\{0,1\}^n$ (see
Lemma \ref{lemma:atomlib}), of $\mathfrak{B}$ can be written uniquely as
$a_\varepsilon=a_{\varepsilon'}\cap a_{\varepsilon''}$ where
$a_{\varepsilon'}$ is an atom of $\mathfrak{B}_1$, $\varepsilon'\in\{0,1\}^k$
and $a_{\varepsilon''}$ is an atom of $\mathfrak{B}_2$,
$\varepsilon'\in\{0,1\}^{n-k}$. Then $\mu$ can be defined as
$\mu(a_\varepsilon)=\mu_1(a_{\varepsilon'})\mu_2(a_{\varepsilon''})$ for all
atoms $a_\varepsilon\in\mathfrak{B}$. It is obviously a measure (non-negative
and additive on disjoint elements); conditions $\mu_{|_{\mathfrak{B}_i}}\equiv
\mu_i$, $i=1,2$ and normalization condition follow easily from
(\ref{eq:atomsubalgebra}) (see Lemma \ref{lemma:atomlib}) and normalization
conditions for $\mu_1$ and $\mu_2$.  $\square$\\

This provides a direct elementary construction of a classical extension 
for a PPT where Boolean subalgebras of
compatible observables have only trivial (i.e. $\{\emptyset,\mathbf{1}\}$)
intersections;this is the case, e.g., of a (finite) 
collection of spin measurements on a spin $1/2$
particle discussed by Bell \cite{Bell66} and Kochen and Specker \cite{K-S}.

\section{\label{sec:ht}Horn-Tarski partial measures}
The notion of \textit{partial measure} was introduced by Horn and Tarski \cite{Tarski} 
to analyze the possibility of extending a measure defined on a
subalgebra of a Boolean algebra, or even a function defined on an arbitrary
subset of the algebra, to a measure  on the entire algebra. Even if such 
a notion was not related to QM by the authors, the reduction of the
extension problem to the case of H-T PPTs transforms the H-T results into
fundamental general criteria of classical representability.  

For the reader's convenience we give therefore
a brief summary of Horn and Tarski's results.

\begin{defin}\label{def:seqord}
 Let $A_0,\ldots,A_{m-1}$ and $B_0,\ldots,B_{n-1}$ be elements of a Boolean
algebra $\mathfrak{B}$, we  say  that
\begin{equation}
 \langle A_0,\ldots,A_{m-1}\rangle\leq \langle B_0,\ldots,B_{n-1}\rangle
\end{equation}
if
\begin{equation}\label{eq:inclus}
 \bigcup_{r\in S_{k,m}}\bigcap_{i\leq k}A_{r_i}\quad\subset\quad \bigcup_{r\in
 S_{k,n}}\bigcap_{i\leq k}B_{r_i}
\end{equation}
for all $k<m$, where $S_{k,n}$ is the set of all sequences of natural numbers
$r=(r_0,\ldots,r_k)$ with $0\leq r_0<\ldots<r_k<n$
\end{defin}

\begin{defin}\label{def:mispar}
A real function $f$ defined over a subset $S$ of a Boolean algebra
$\mathfrak{B}$ is  a \textit{partial measure} if the following conditions are
satisfied
\begin{itemize}
 \item[(i)] $f(x)\geq 0$ for all $x\in S$.
\item[(ii)] If $A_0,\ldots,A_{m-1}$ and $B_0,\ldots,B_{n-1}$ are elements of
$S$ and
\begin{displaymath}
 \langle A_0,\ldots,A_{m-1}\rangle\leq \langle B_0,\ldots,B_{n-1}\rangle,
\end{displaymath}
 then
\begin{displaymath}
 \sum_{i<m} f(A_i)\leq \sum_{j<n} f(B_j).
\end{displaymath}
\item[(iii)] $\textbf{1}\in S$ and $f(\textbf{1})=1$.
\end{itemize}

\end{defin}

An obvious consequence of Definition \ref{def:mispar} is that if
$\textbf{1}\in S\subset T$ and $f$ is a partial measure on $T$, then $f$ is a
partial measure on $S$. Another important property is shown by the following
\begin{teo}\label{teo:subalg}
 Let $S\subset \mathfrak{B}$ be a subalgebra of $\mathfrak{B}$ (in particular
 $S=\mathfrak{B}$); then a function  $f$ on $S$ is a partial measure on $S$
 $\Longleftrightarrow$ $f$ is a measure on $S$.
\end{teo}

As shown by Theorem \ref{teo:feg}, a fundamental role is played by the
following notions

\begin{defin}\label{def:intest}
 Let $S\subset \mathfrak{B}$, and $f$ be a partial measure on $S$ and $x\in
 \mathfrak{B}$. We define the \textit{exterior measure} of $x$ with respect to
 $f$, and we write $f_e(x)$, as the greatest lower bound of numbers $\xi$ of
 the form
\begin{equation}\label{eq:xi}
 \xi=\frac{1}{m}\left[ \sum_{i<n} f(A_i)-\sum_{j<p} f(B_j)\right]
\end{equation}
where $A_i,B_j\in S$ for $i<n,j<p$ and where
\begin{equation}
 \langle B_0,\ldots,B_{p-1},x_0,\ldots,x_{m-1}\rangle \leq\langle
 A_0,\ldots,A_{n-1}\rangle.
\end{equation}
with $x_i=x$ for all $i<m$.

Similarly, we define the \textit{interior measure} of $x$ with respect to $f$,
and we write $f_i(x)$, as the least upper bound of numbers $\xi$ of the form
(\ref{eq:xi}), where
\begin{equation}
 \langle A_0,\ldots,A_{n-1}\rangle \leq \langle
 B_0,\ldots,B_{p-1},x_0,\ldots,x_{m-1}\rangle.
\end{equation}
with $x_i=x$ for all $i<m$.
\end{defin}

\begin{teo}\label{teo:intest}
 If $f$ is a (partial) measure on a subalgebra $\mathfrak{B}_0\subset
 \mathfrak{B}$, then \\$f_i(x)=\sup\{f(y)|y\in \mathfrak{B}_0 , y\subset x\}$
 and $f_e(x)=inf\{f(y)| y\in\mathfrak{B}_0, x\subset y\}$.
\end{teo}

Horn and Tarski's main results are the following

\begin{teo}\label{teo:feg}
 Let $f$ be a partial measure on $S\subset\mathfrak{B}$, $x\in\mathfrak{B}$
and $g$ be a function on $S\cup\{x\}$  that coincide with $f$ on $S$; then $g$
is a partial measure on $S\cup\{x\}$ $\Longleftrightarrow$ ${f_i(x)\leq
g(x)\leq f_e(x)}$.
\end{teo}

\begin{teo}\label{teo:sint}
 Let $f$ be a partial measure on $S$ and $S\subset T\subset \mathfrak{B}$;
then a partial measure $g$ on $T$ that  coincide with $f$ on $S$ exists.
\end{teo}

\begin{teo}\label{teo:estmisparz}
 Let $f$ be a partial measure on a subset $S$ of a Boolean algebra
$\mathfrak{B}$; then a measure $\mu$ on  $\mathfrak{B}$ that coincide with $f$
on $S$ exists.
\end{teo}

We remark that Horn and Tarski's notion of partial measure has nothing to do
with partial Boolean algebras, in the Kochen-Specker or in our version, 
nor with partial probability theories;
even if H-T partial measures are defined only on subsets, they satisfy 
conditions which imply their
\textit{extensibility to measures on the entire algebra}.
Our approach is rather based on a comparison of the two notions, 
with the result of transforming the H-T conditions for H-T partial measures 
into extensibility conditions for PPTs, in the case of PBAs which are embeddable 
in a Boolean algebra. 

Since the condition in Theorem \ref{teo:estmisparz} is obviously also
necessary, the extensibility criterion provided by partial measures is
equivalent to the correlation polytope criterion. 
In particular, this implies that inequalities derived from the correlation polytope
criterion (Proposition \ref{prop:polit}) are equivalent to those derived from
condition $(ii)$ of definition \ref{def:mispar}. 

It should be also remarked that, while Pitowsky's correlation polytopes 
implicitly assume a free Boolean structure, 
Horn and Tarski's extensibility criterion also applies to non-free 
Boolean algebras and to more general subsets with respect to
those considered in Proposition 
\ref{prop:polit}; in principle, therefore, the H-T criterion also allows
for a direct analysis of extensions of states on a PBA in the 
framework of projection algebra PPTs.

\end{document}